\documentclass[a4paper,fleqn,numbers]{cas-sc}

\usepackage{iftex}
\ifPDFTeX
  \usepackage[utf8]{inputenc}
  \usepackage[T1]{fontenc}
  \usepackage{graphicx}
  \usepackage{hyperref}
\else\ifXeTeX
  \usepackage{graphicx}
  \usepackage{hyperref}
\else\ifLuaTeX
  \usepackage{graphicx}
  \usepackage{hyperref}
\else
  \usepackage[dvipdfmx]{graphicx}
  \usepackage[dvipdfmx]{hyperref}
\fi
\fi
\fi
\usepackage{amsmath}
\usepackage{amssymb}
\usepackage{url}
\usepackage{booktabs}
\usepackage{caption}
\usepackage{subcaption}
\usepackage{indentfirst}
\usepackage{makecell}
\usepackage[numbers,sort&compress]{natbib}

\makeatletter

\def\@journal{}
\makeatother

\begin{document}
\let\WriteBookmarks\relax
\def\floatpagepagefraction{1}
\def\textpagefraction{.001}
\setcounter{secnumdepth}{-1}

\shorttitle{}
\shortauthors{}

\title[mode=title]{Dynamic Kinetic Evaluation Favors Compositionally Diverse Multicomponent Alloy Nanoparticles for Hydrogen Evolution}

\author[1]{Koki Otsuka}
\credit{Conceptualization, Methodology, Software, Investigation, Formal analysis, Writing - original draft}

\author[1,2]{Anh Khoa Augustin Lu}
\credit{Writing - review and editing}

\author[3]{Koji Shimizu}
\credit{Investigation, Data curation}

\author[1]{Satoshi Watanabe}
\cormark[1]
\ead{watanabe@cello.t.u-tokyo.ac.jp}
\credit{Writing - review and editing, Supervision}

\affiliation[1]{organization={Department of Materials Engineering, The University of Tokyo},
            postcode={113-8656},
            city={Tokyo},
            country={Japan}}

\affiliation[2]{organization={Research Center for Materials Nanoarchitectonics, National Institute for Materials Science},
            postcode={305-0044},
            city={Tsukuba},
            country={Japan}}

\affiliation[3]{organization={Materials DX Research Center, National Institute of Advanced Industrial Science and Technology},
            postcode={305-8568},
            city={Tsukuba},
            country={Japan}}

\cortext[1]{Corresponding author}

\begin{graphicalabstract}
\centering
\includegraphics[width=\textwidth]{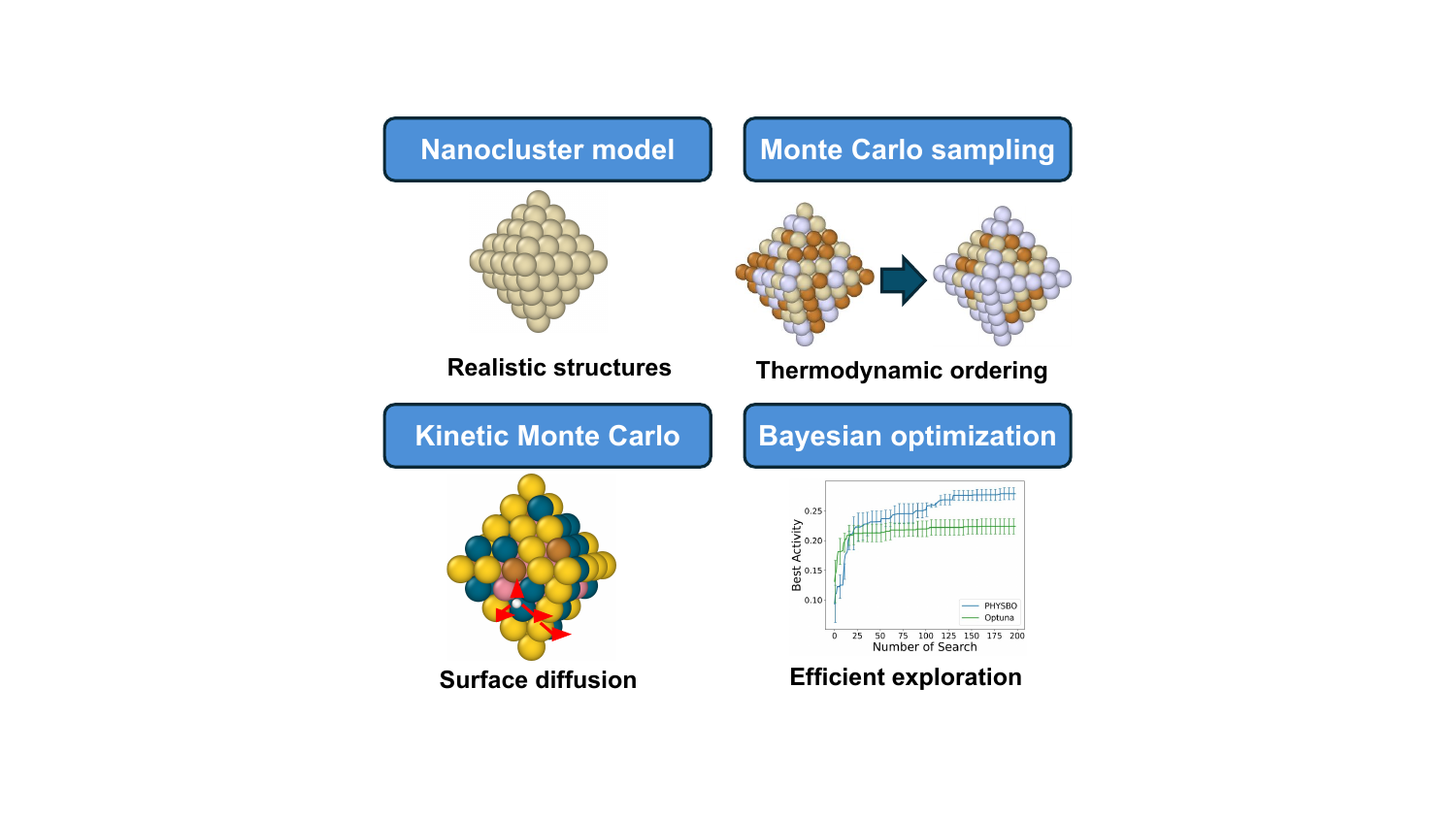}
\end{graphicalabstract}

\begin{abstract}
High-entropy alloy (HEA) nanoparticles offer chemically diverse catalytic sites, but identifying optimal compositions across their large design space remains challenging. We developed a framework combining machine-learning energy prediction, thermodynamic Monte Carlo sampling, Bayesian optimization, and kinetic Monte Carlo (kMC) simulations to screen 85-atom nanoparticle models across a nine-element composition space for the hydrogen evolution reaction. Static optimization based on adsorption energies selected Pt-rich, low-component compositions, whereas kMC-based optimization ranked the multielement composition Au$_{33}$Co$_{20}$Cu$_{6}$Pd$_{26}$ highest among the evaluated candidates. Cycle-resolved analysis further showed hydrogen adsorption at Co/Pd-rich environments, migration through intermediate-binding sites, and frequent Heyrovsky desorption at weaker-binding, Au-containing environments. This pathway-level evidence supports a multisite reaction picture and illustrates how incorporating kinetics can complement static adsorption-energy screening.
\end{abstract}


\maketitle

\section{Introduction}
High-entropy alloys (HEAs) are broadly defined as multicomponent alloys composed of several principal elements at non-dilute concentrations, often in near-equiatomic ratios. In contrast to conventional alloy design, which is typically centered on a dominant host element, HEAs span compositionally complex design spaces in which high configurational entropy can promote the formation of single-phase solid solutions, while local lattice distortion and chemically diverse nearest-neighbor environments give rise to unusual structure--property relationships \cite{Miracle2017,George2019}. These characteristics have attracted growing interest not only in structural materials science but also in catalysis, where atomic-scale chemical heterogeneity creates diverse active-site environments and reaction energetics, providing a broad design space for catalytic functionality.

This broad design space offers opportunities to overcome limitations associated with monometallic and simple bimetallic catalysts. In HEAs, distinct local elemental motifs generate near-continuous distributions of adsorption energies and activation barriers, enabling flexible tuning of catalytic activity and selectivity \cite{Batchelor2019,Xin2020}. This concept has already been explored across multiple electrocatalytic reactions, including the oxygen reduction reaction (ORR) \cite{Chen2023ORR} and carbon dioxide/carbon monoxide reduction (CO$_2$RR/CORR), through both computational screening and experimental studies \cite{Pedersen2020CO2,Nellaiappan2020}. At the same time, rapid progress in synthetic chemistry has enabled the fabrication of ultrasmall HEA nanoparticles with controlled compositions, further expanding their practical relevance to electrochemical energy conversion \cite{Mahin2025}.

Among these applications, the hydrogen evolution reaction (HER) is a particularly important target because efficient and sustainable hydrogen production is essential for the transition to low-carbon energy systems. Recent studies have reported promising HER performance in HEA nanoparticles, including sub-2 nm systems and compositions containing platinum-group metals, together with high catalytic stability \cite{Wu2020,Feng2021}. These findings indicate that HEA nanoparticles can combine high intrinsic activity with structural robustness, supporting their potential as HER electrocatalysts \cite{Sun2024,Shi2024}.

However, identifying optimal HEA compositions is a high-dimensional optimization problem, and exhaustive first-principles screening is generally impractical. Data-driven approaches combining machine-learning (ML) models with efficient optimization algorithms, such as Bayesian optimization, are therefore increasingly used to navigate broad compositional spaces with substantially reduced computational cost \cite{Pedersen2021}.

Despite these advances, several major methodological challenges remain in the rational design of HEA catalysts. First, the accessible compositional space grows combinatorially with the number of constituent elements, whereas many established computational and experimental optimization studies have focused on predefined composition spaces containing approximately five elements \cite{Batchelor_2021,Pedersen2020CO2,Pedersen2021}. Although more recent approaches have extended Bayesian optimization to substantially larger spaces comprising up to ten elements \cite{Xu2024}, efficient exploration of the vast elemental and compositional space remains a fundamental challenge. Second, many computational screening frameworks rely on idealized periodic slab models, often with a limited number of surface orientations \cite{Pedersen2020CO2,Wang_2025}. Such models cannot fully capture the geometric complexity of experimentally relevant nanoparticles, including finite-size effects and the presence of edges, corners, and other low-coordination sites, although recent studies have begun to address HEA nanoparticles explicitly \cite{Shiota_2025}. Third, screening calculations frequently assume random or otherwise prescribed atomic arrangements and do not explicitly account for finite-temperature thermodynamic effects such as surface segregation and chemical ordering, which can substantially alter the surface composition and local atomic environments \cite{Kristoffersen_2022,Ye_2026}. For example, Pedersen et al. experimentally observed two-phase coexistence in an Ir--Pt composition spread, revealing incomplete mixing that was not represented in the idealized solid-solution model used for activity optimization \cite{Pedersen2021}. Such phase separation, segregation, and local ordering can therefore contribute to deviations between computationally predicted and experimentally realized catalyst structures and activities.

Finally, most conventional high-throughput screening strategies evaluate catalytic performance primarily through the static adsorption energetics of individual local environments, often within Sabatier- or volcano-type frameworks \cite{Pedersen2020CO2,Wang_2025,Rittiruam_2024}. While highly successful for identifying favorable binding environments, such approaches may overlook cooperative mechanisms involving multiple chemically distinct sites and the transport of reaction intermediates between them. Recent work has demonstrated that HEAs can exploit such multisite mechanisms. For hydrogen evolution on PtFeCoNiCu, for example, Chen et al. showed theoretically that strong hydrogen-binding sites can facilitate hydrogen adsorption, while adsorbed hydrogen can diffuse to weaker-binding sites that favor H$_2$ formation and desorption, thereby enabling catalytic behavior that cannot be described by the conventional single-site Sabatier picture \cite{Chen2024}. These findings suggest that optimization based solely on static, independently evaluated adsorption sites may fail to identify compositions that benefit from cooperative reaction dynamics, motivating approaches that explicitly consider both the structural complexity of HEA nanoparticles and diffusion-mediated kinetic coupling between chemically and functionally distinct active sites.

To address these challenges, we develop a composition-optimization framework for multicomponent alloy nanoparticle catalysts for HER that integrates ML-based energy prediction, Monte Carlo (MC) sampling of thermodynamically accessible atomic configurations, data-driven composition-space exploration, and kinetic Monte Carlo (kMC) simulations of hydrogen adsorption, surface diffusion, and desorption. Using an 85-atom octahedral nanoparticle model spanning a nine-element chemical space, we compare static adsorption-energy-based screening with a kinetics-aware evaluation of adsorbate transport across heterogeneous surface sites. The search allows the number and proportions of constituent elements to vary, rather than restricting candidates to near-equiatomic HEAs. The two evaluations select compositionally distinct candidates, and ensemble trajectory analysis is used to examine diffusion-mediated coupling among chemically distinct sites in the kMC-selected composition.

\section{Results}

\subsection{Static Screening Results}

Because the elemental distribution within multicomponent alloy nanoparticles is not random but is governed by thermodynamic preferences for surface segregation and local ordering, we generated representative atomic configurations for each composition using Metropolis Monte Carlo (MC) sampling at 900~K (see Methods for details). As a static activity metric, we defined $A$ as the arithmetic average of Boltzmann-type site-specific activity scores across all adsorption sites of these thermodynamically sampled atomic configurations. Each site score follows the Sabatier principle: a site is most active when the ML-predicted hydrogen adsorption energy $\Delta E_{\mathrm{H}}$ is close to $\Delta E_{\mathrm{opt}}=-0.27$~eV, which corresponds to $\Delta G_{\mathrm{H}^*}\approx-0.03$~eV under the $+0.24$~eV correction used in this study, and the score decays exponentially with deviation from this optimum (see Methods for full details). Using $A$ as the objective function, composition optimization was performed using two approaches: PHYSBO\cite{Motoyama2022}, a candidate-based Bayesian optimization library, and Optuna\cite{Akiba2019}, a Tree-structured Parzen Estimator (TPE) framework, as described in the Methods section.

\subsubsection{Optimization Trajectories and Discovered Compositions}
\begin{figure}[pos=h]
    \centering
    \includegraphics[width=0.55\textwidth]{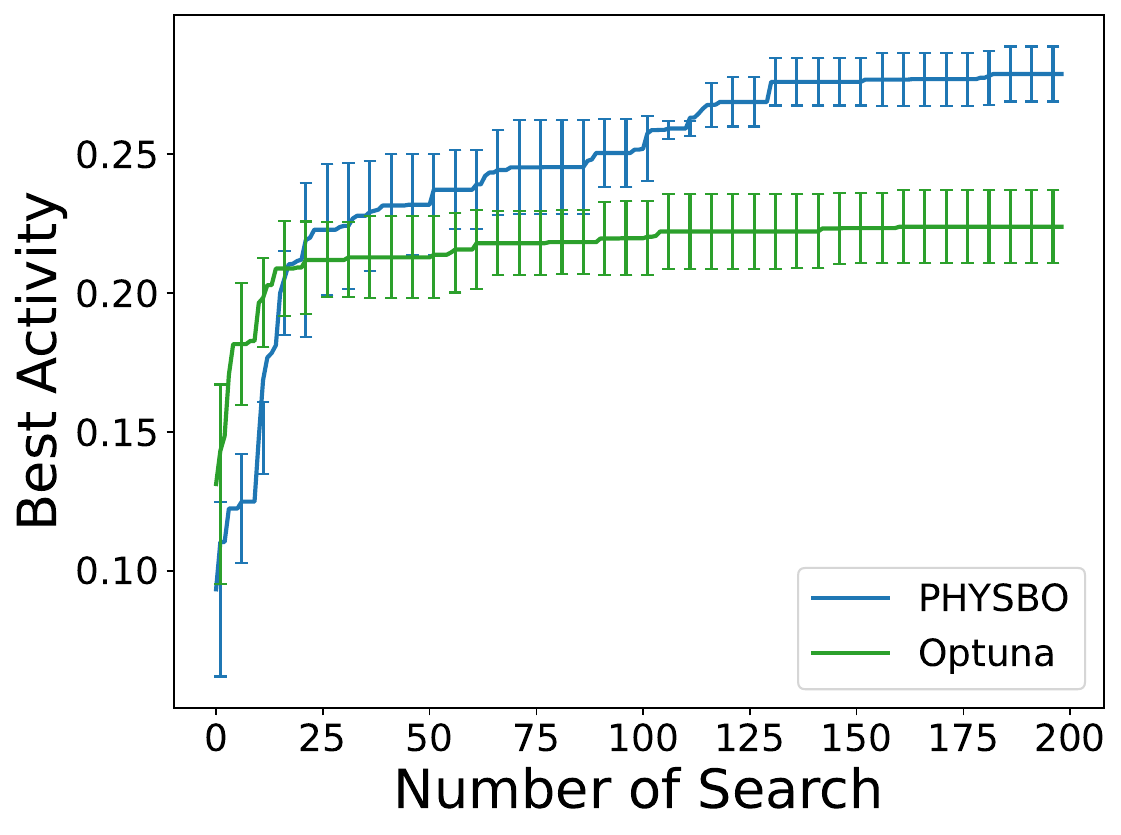}
    \caption{Optimization trajectories of the HER activity obtained using PHYSBO and Optuna. The plot shows the best objective-function value found up to each iteration. The 200-iteration search was independently repeated five times for each method, with the solid lines and error bars representing the mean and standard deviation across the runs, respectively. PHYSBO reached higher best-so-far activity scores than Optuna.}
    \label{fig:opt_trajectory}
\end{figure}

Figure \ref{fig:opt_trajectory} shows the optimization trajectories in terms of the best-so-far activity score obtained up to each iteration. To evaluate variation among independent searches, 200 optimization iterations were performed five times for each method. The solid lines represent the mean best-so-far activity across the five runs, and the error bars represent the standard deviation across the runs. Both methods identified compositions with higher activity scores than those found during the initial random-search stage in iterations 0 to 10. PHYSBO showed stepwise improvements from the early to intermediate stages of the search and reached higher final activity scores on average than Optuna. By contrast, the mean best-so-far activity obtained with Optuna plateaued at a lower value. Moreover, the highest-ranking Optuna candidates in the best-performing run were all Pd-Pt binary compositions, indicating that the search was concentrated in a relatively narrow compositional region.

Table \ref{tab:her_best_compositions} presents the five highest-activity compositions identified within the best-performing run for each method. The highest activity score obtained with PHYSBO was 0.2894 for Pd$_{10}$Pt$_{58}$Rh$_{17}$. The highest-ranking PHYSBO compositions were predominantly ternary alloys containing Pt as the major component, with smaller amounts of Pd and Rh. Relative to the pure Pt cluster, for which $A \approx 0.20$, Pd$_{10}$Pt$_{58}$Rh$_{17}$ showed an approximately 1.45-fold higher static activity score. This result suggests that alloying Pt with Pd and Rh can produce local adsorption environments that are more favorable under the present static activity metric.

\begin{table}[pos=h]
    \centering
    \caption{Five highest-activity compositions identified within the best-performing run for each optimization method.}
    \label{tab:her_best_compositions}
    \begin{tabular}{|l|c|l|c|}
    \hline
    \multicolumn{2}{|c|}{PHYSBO}&\multicolumn{2}{c|}{Optuna}\\
    \hline
    Composition & $A$ & Composition & $A$\\ \hline
    Pd$_{10}$Pt$_{58}$Rh$_{17}$ & 0.2894 & Pd$_{29}$Pt$_{56}$ & 0.2639 \\
    Pd$_{7}$Pt$_{65}$Rh$_{13}$ & 0.2893 & Pd$_{27}$Pt$_{58}$ & 0.2612 \\
    Pd$_{8}$Pt$_{64}$Rh$_{13}$ & 0.2862 & Pd$_{28}$Pt$_{57}$ & 0.2609 \\
    Pd$_{7}$Pt$_{67}$Rh$_{11}$ & 0.2846 & Pd$_{30}$Pt$_{55}$ & 0.2608 \\
    Pd$_{6}$Pt$_{64}$Rh$_{15}$ & 0.2840 & Pd$_{31}$Pt$_{54}$ & 0.2582 \\
    \hline
    \end{tabular}
\end{table}

\subsubsection{Site-Level Activity Distribution and Core-Shell Ordering}
\begin{figure}[pos=h]
    \centering
    \includegraphics[width=0.25\textwidth]{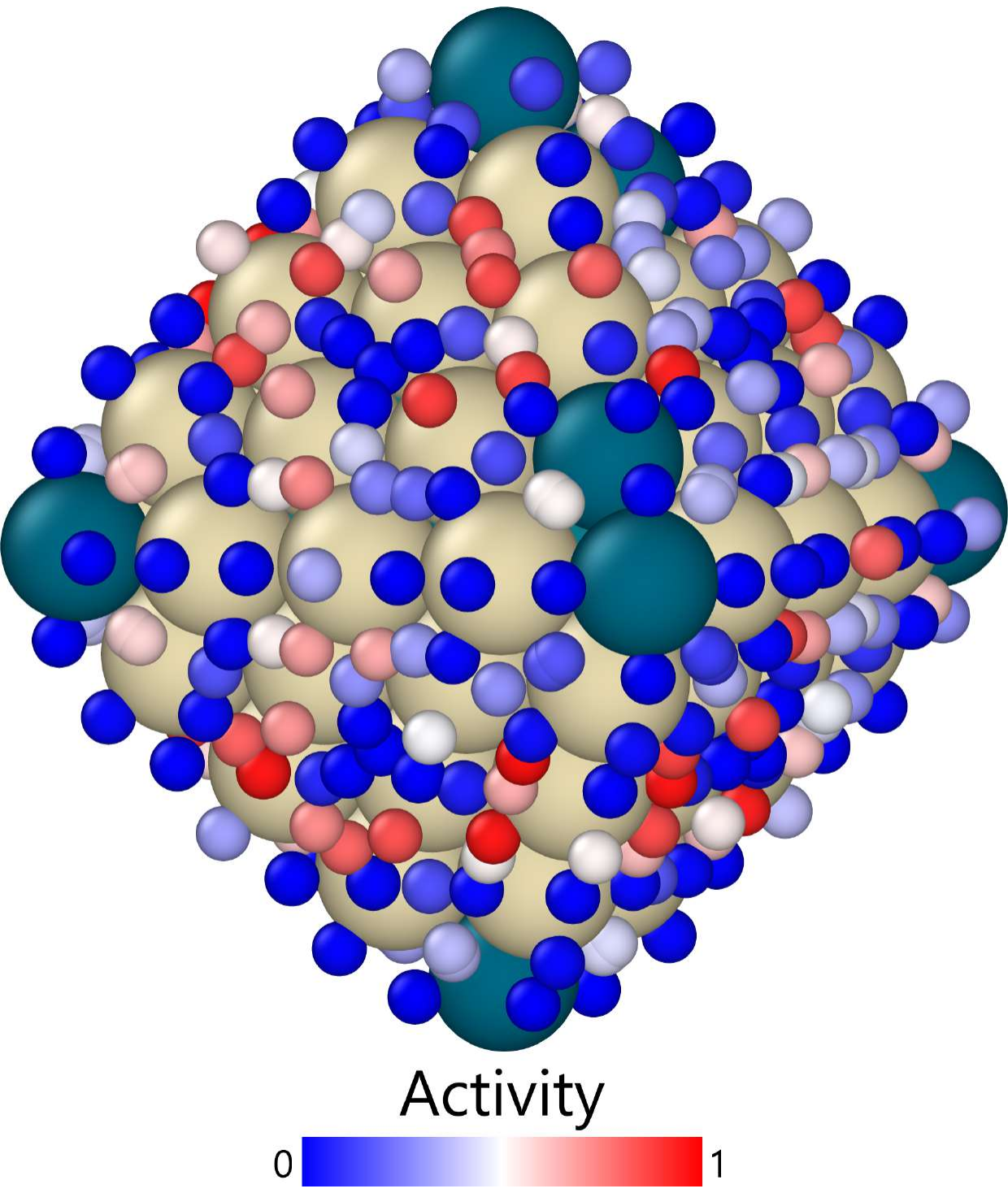} 
    \caption{Spatial distribution of HER active sites on the optimized Pd$_{10}$Pt$_{58}$Rh$_{17}$ cluster. The large spheres represent metal atoms (beige: Pt, dark blue: Pd), and the small spheres indicate the hydrogen adsorption sites (Rh atoms are located inside the cluster). The color of the sites corresponds to the local activity score, with red indicating high activity.}
    \label{fig:spatial_distribution}
\end{figure}

To elucidate the origin of the high static activity score of the best composition identified by PHYSBO, Pd$_{10}$Pt$_{58}$Rh$_{17}$, we visualized the spatial distribution of the local activity scores on the cluster surface using an atomic configuration sampled by MC. Figure \ref{fig:spatial_distribution} shows this distribution. The large spheres represent the metal atoms (beige: Pt and dark blue: Pd), while the small spheres above the surface indicate the adsorption sites, including on-top, bridge, and hollow sites. Rh atoms are located primarily in the cluster interior and are therefore not visible in this surface representation. The site colors represent the local activity scores, with red indicating higher values.

Analysis of the MC-sampled atomic configurations revealed a tendency toward Rh-core/Pt-shell-like ordering rather than a homogeneous elemental distribution, with Pt atoms forming most of the surface shell and Pd preferentially occupying sites near the vertices. Rh atoms were located primarily in the cluster interior, while Pt and Pd were more prevalent at the surface.

The surface Pt sites in this Rh-core/Pt-shell-like configuration include sites with high local activity scores, indicating that Pt at the surface of an Rh-rich nanoparticle can provide favorable hydrogen adsorption environments according to the present static activity metric. A related motif was reported by Guo et al.\cite{Guo2022}, who synthesized Rh@Pt core-shell nanoparticles and attributed their enhanced HER activity relative to pure Pt to changes in hydrogen adsorption on the Pt shell induced by compressive strain from the Rh core. This experimental result is consistent with the high activity associated with Pt over an Rh-rich interior in our model, although the present calculations do not isolate or quantify a strain contribution.

\subsection{Dynamic kMC Evaluation Results}

To capture the role of surface diffusion in activity, we employed kinetic Monte Carlo (kMC) simulations in which single hydrogen atoms undergo three elementary processes on the nanoparticle surface: (i) Volmer adsorption, (ii) surface diffusion between neighboring sites, and (iii) Heyrovsky desorption as H$_2$. For each thermodynamically sampled atomic configuration, the H$_2$ production rate was calculated as $r=N_{\mathrm{H}_2}/t_{\mathrm{total}}$. The dynamic activity score $A_{\mathrm{kMC}}$ is the natural logarithm of the configurationally averaged rate (see Methods for full details); therefore, a larger (less negative) value indicates a higher predicted production rate. Using $A_{\mathrm{kMC}}$ as the objective function, the compositional space was explored using PHYSBO over 200 iterations. The optimization identified high-activity compositions from the early stages of the search, with Au$_{33}$Co$_{20}$Cu$_{6}$Pd$_{26}$ exhibiting the highest $A_{\mathrm{kMC}}$ among the evaluated compositions. Whereas static screening based on adsorption energies favored Pt- or Rh-rich compositions containing relatively few elements, the kMC-based evaluation favored multielement compositions with chemically diverse local environments. Table \ref{tab:kmc_top5} lists the five compositions with the highest $A_{\mathrm{kMC}}$ values identified during the optimization.

\begin{table}[pos=h]
    \centering
    \caption{Five compositions with the highest kMC activity scores, $A_{\mathrm{kMC}}=\ln(\langle r\rangle)$, identified by the kMC-based optimization.}
    \begin{tabular}{|l|c|}
        \hline
        Composition & $A_{\mathrm{kMC}}$\\ \hline
        Au$_{33}$Co$_{20}$Cu$_{6}$Pd$_{26}$ & -18.723\\
        Au$_{30}$Co$_{21}$Pd$_{34}$ & -18.757 \\
        Ag$_{1}$Au$_{31}$Co$_{18}$Cu$_{7}$Pd$_{28}$ & -18.781\\
        Au$_{33}$Co$_{15}$Cu$_{12}$Pd$_{25}$ & -18.864\\
        Au$_{33}$Co$_{16}$Pd$_{36}$ & -18.885\\
        \hline
    \end{tabular}
    \label{tab:kmc_top5}
\end{table}

\begin{figure}[pos=h]
    \centering
    \includegraphics[width=0.25\textwidth]{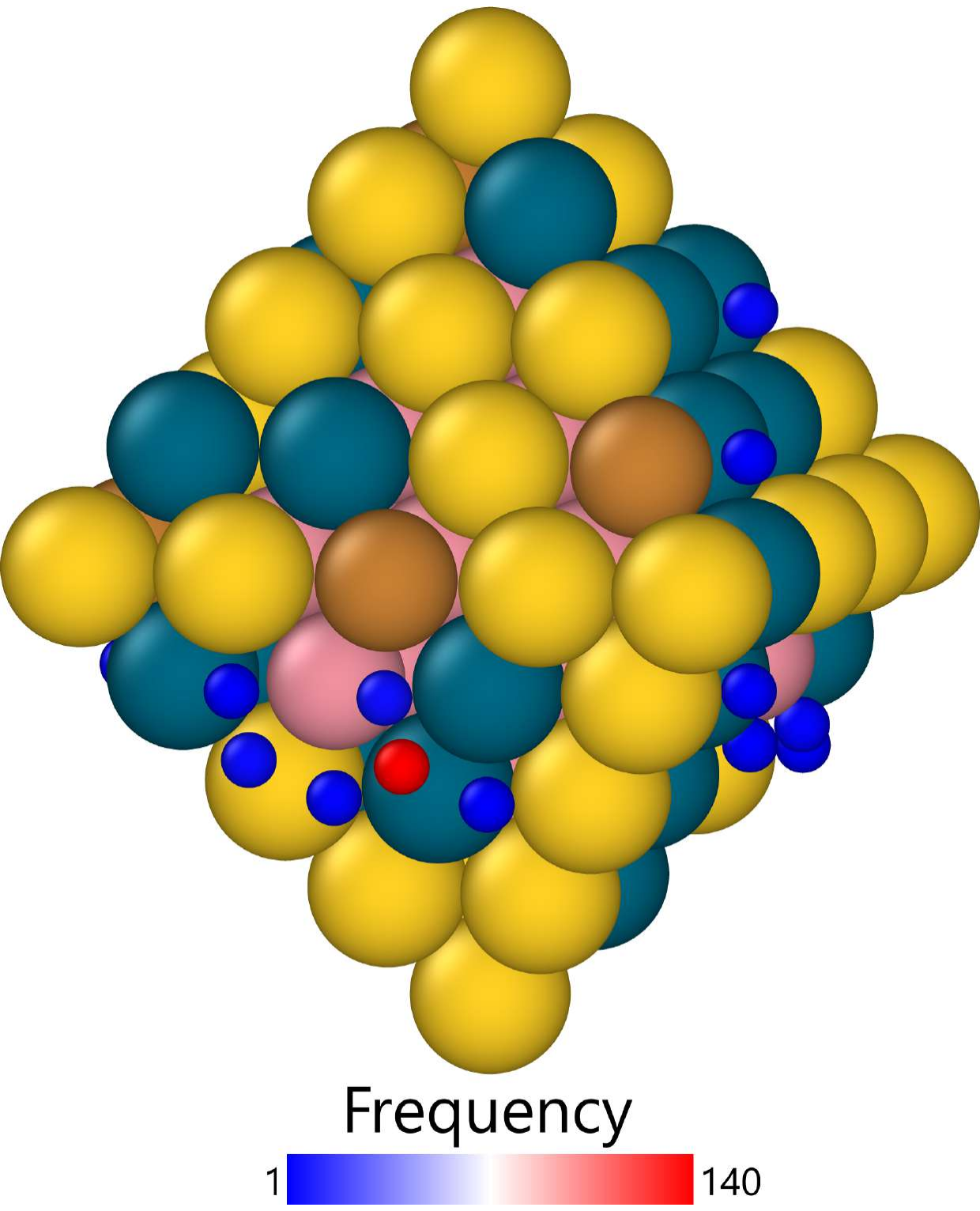}
    \includegraphics[width=0.25\textwidth]{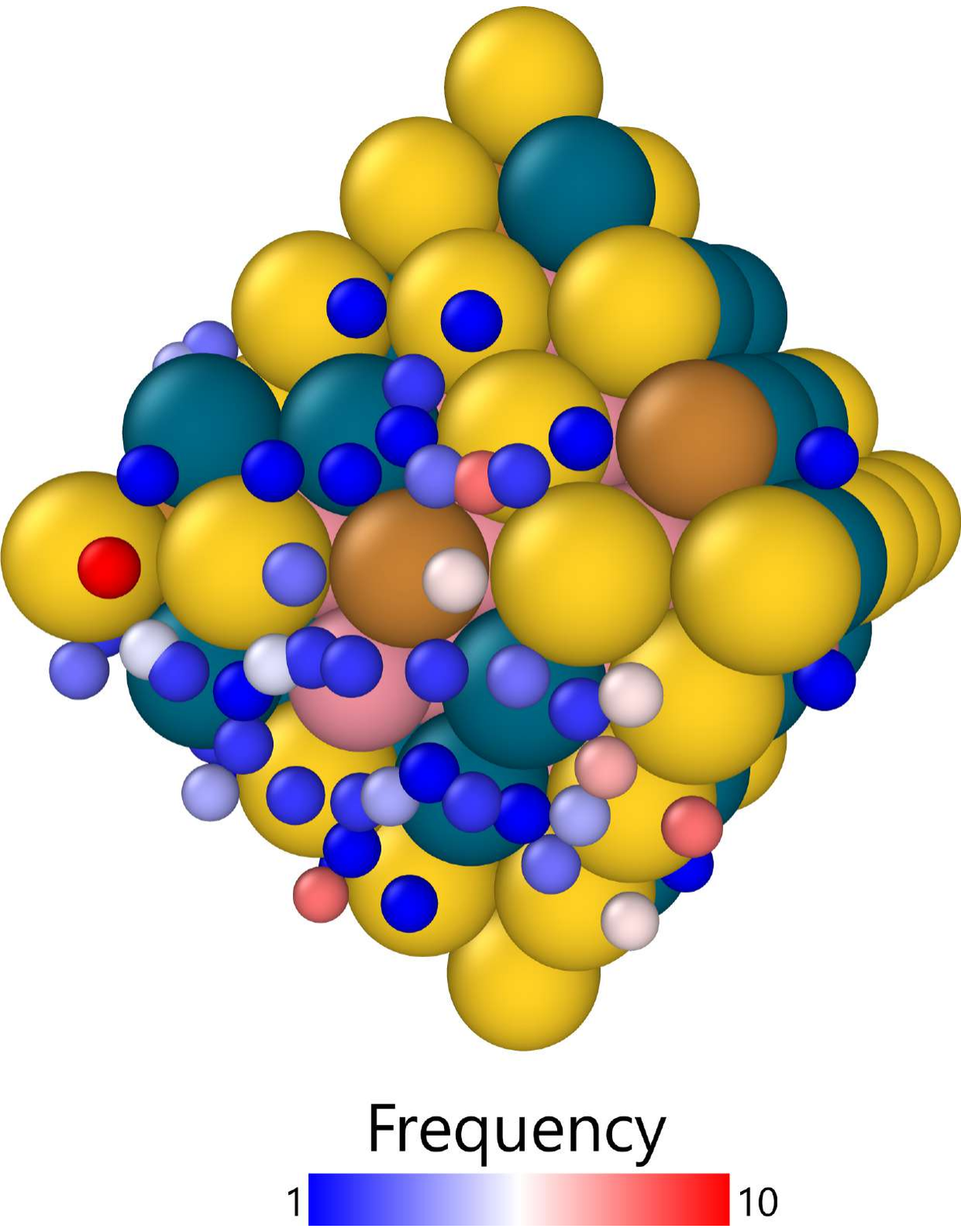}
    \caption{Spatial distributions of kMC event frequencies on the Au$_{33}$Co$_{20}$Cu$_{6}$Pd$_{26}$ nanoparticle surface. (a) Hydrogen adsorption frequency. (b) $\mathrm{H}_2$-formation and desorption frequency. The metal atoms are represented by Au in yellow, Pd in blue, Cu in brown, and Co in pink. Darker red adsorption-site markers indicate higher event frequencies. Sites with zero recorded events are not shown.}
    \label{fig:kmc_events}
\end{figure}

To investigate the origin of the high dynamic activity of Au$_{33}$Co$_{20}$Cu$_{6}$Pd$_{26}$, we first analyzed the spatial distributions of adsorption and desorption events on the nanoparticle surface. As shown in Figure \ref{fig:kmc_events}, hydrogen adsorption events occurred predominantly at hollow sites surrounded by strong hydrogen-binding elements, particularly Pd and Co. In contrast, $\mathrm{H}_2$-formation and desorption events occurred frequently at more weakly binding sites, including Au-containing sites and Au-Pd interfacial environments. These spatially distinct event distributions are consistent with a cooperative pathway in which hydrogen adsorbs at strong-binding sites, migrates across the surface, and undergoes Heyrovsky desorption at weaker-binding sites.

\begin{figure}[pos=h]
    \centering
    \includegraphics[width=\textwidth]{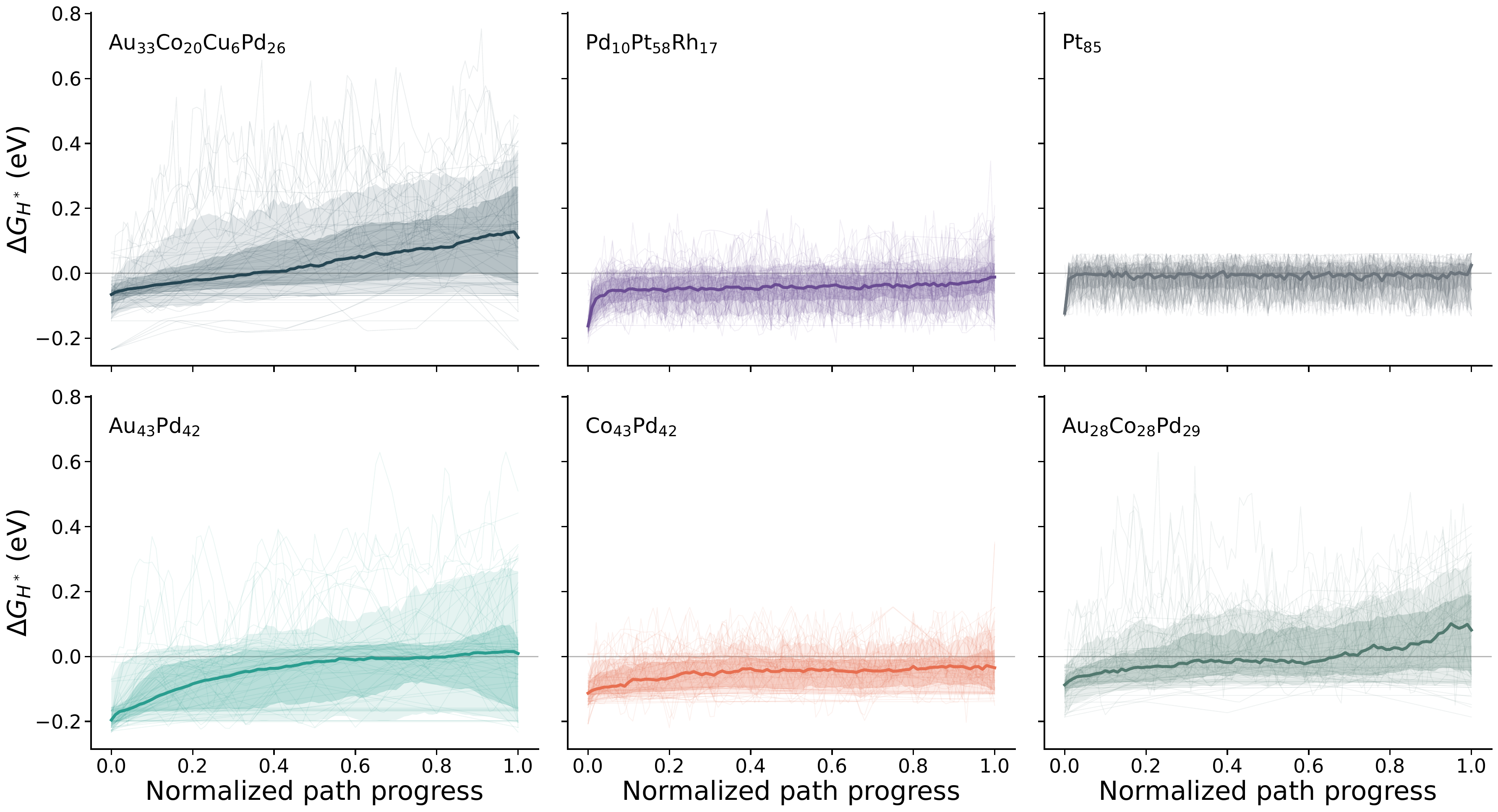}
    \caption{Ensemble path-energy profiles for $\mathrm{H}_2$-production cycles in kMC simulations. For each cycle, the horizontal axis gives relative progress through the initial adsorption site, the sites in first-visit order, and the site of the final Heyrovsky event. A desorption site visited earlier in the cycle is repeated at the endpoint so that 1 always represents the actual desorption event. Sequences with different numbers of sites are linearly interpolated onto this shared coordinate; it does not represent elapsed time or physical distance. Thin lines indicate a random subset of up to 70 individual cycles per composition. The thick line and shaded regions summarize all analyzed cycles and indicate the pointwise median, 25th--75th percentile range, and 10th--90th percentile range, respectively. The plotted free energy is evaluated as $\Delta G_{\mathrm{H}^*}=\Delta E_{\mathrm{H}}+0.24$~eV. The horizontal line at 0~eV marks thermoneutral H adsorption.}
    \label{fig:path_profiles}
\end{figure}

\begin{figure}[pos=h]
    \centering
    \includegraphics[width=\textwidth]{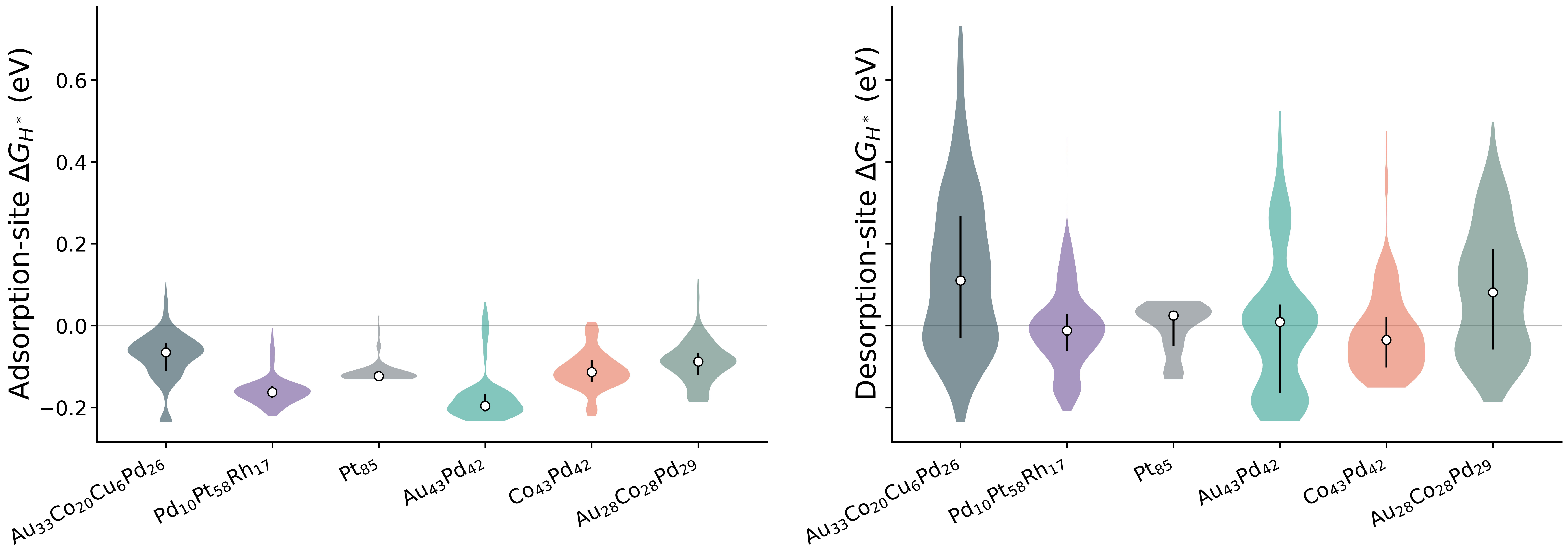}
    \caption{Adsorption and desorption free-energy distributions obtained from the kMC trajectory ensembles for the six selected compositions. For each completed $\mathrm{H}_2$-production cycle, the initial Volmer adsorption site and final Heyrovsky desorption site were identified, and their $\Delta G_{\mathrm{H}^*}$ values were compared. The violin width represents the probability density, while the white circles and black lines indicate the median and interquartile range, respectively. The displacement between the paired distributions visualizes the extent to which adsorption and $\mathrm{H}_2$ desorption sample energetically distinct surface environments under the present single-adsorbate model.}
    \label{fig:ads_des_energy_distribution}
\end{figure}

\begin{figure}[pos=h]
    \centering
    \includegraphics[width=0.58\textwidth]{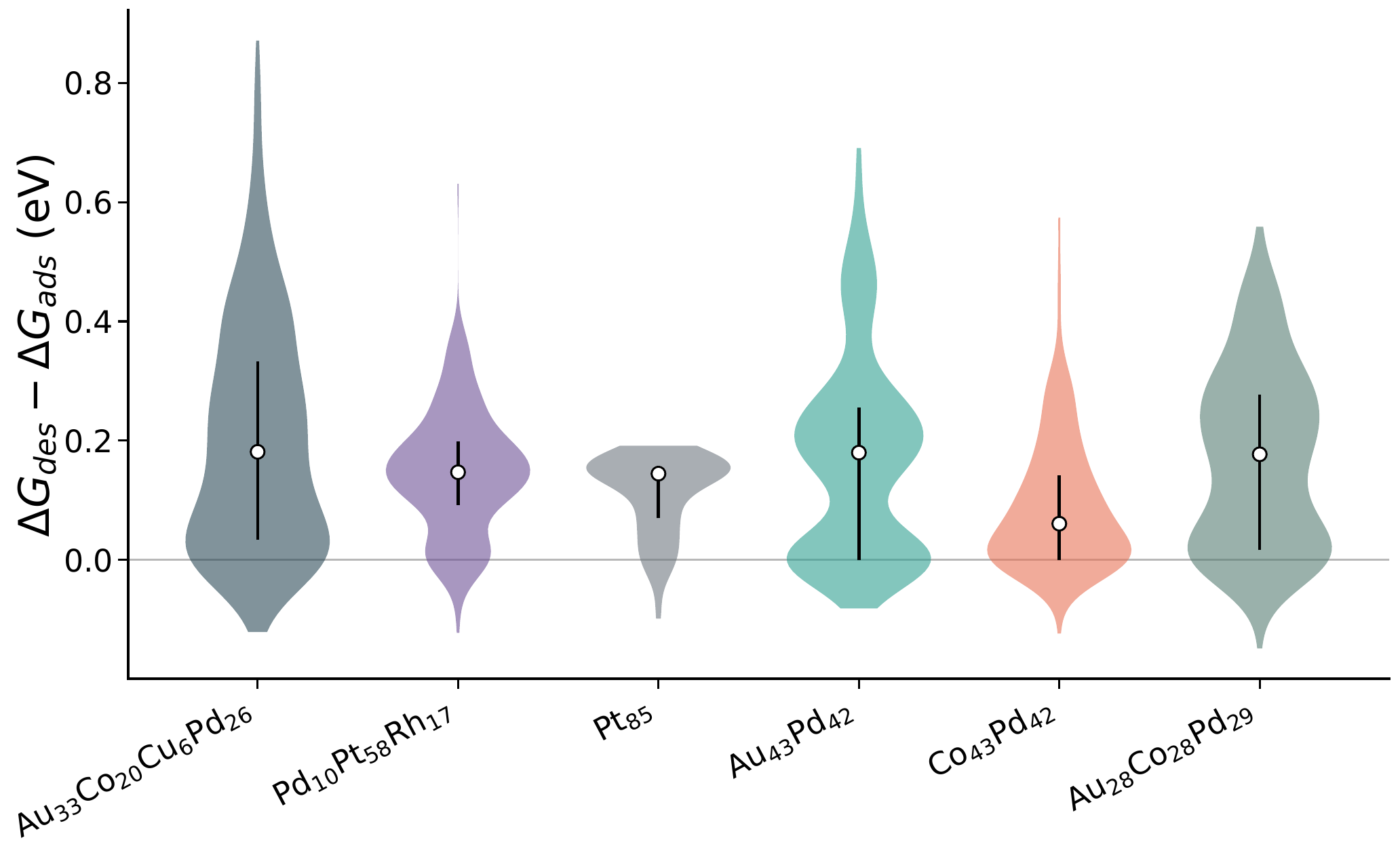}
    \caption{Cycle-wise free-energy shifts extracted from the same kMC trajectory ensembles as in Figure \ref{fig:ads_des_energy_distribution}. The plotted quantity is the difference between the final Heyrovsky desorption site and the initial Volmer adsorption site, $\Delta G_{\mathrm{des}}-\Delta G_{\mathrm{ads}}$, for each completed $\mathrm{H}_2$-production cycle. The violin width represents the probability density, while the white circles and black lines indicate the median and interquartile range, respectively. Positive values indicate that desorption occurs at a weaker-binding site than the site where hydrogen initially adsorbed.}
    \label{fig:energy_shift_range}
\end{figure}

The event-frequency maps identify spatial specialization of the surface sites, but they do not show how the binding environment evolves within individual reaction cycles or whether the same behavior persists across an ensemble of trajectories. To address these questions, the kMC procedure used to evaluate $A_{\mathrm{kMC}}$ was extended to record the adsorption free energies of the sites sampled during each completed $\mathrm{H}_2$-production cycle, from the initial Volmer adsorption event through surface diffusion to the final Heyrovsky event. The site energies were evaluated as $\Delta G_{\mathrm{H}^*}=\Delta E_{\mathrm{H}}+0.24$~eV. We compared six compositions selected to facilitate comparison of the two optimization outcomes and to provide references for the elemental motifs present in the kMC-selected composition. Au$_{33}$Co$_{20}$Cu$_6$Pd$_{26}$ is the kMC optimum, Pd$_{10}$Pt$_{58}$Rh$_{17}$ is the static-screening optimum, and Pt$_{85}$ serves as a monometallic reference. The binary controls Au$_{43}$Pd$_{42}$ and Co$_{43}$Pd$_{42}$ probe Au/Pd and Co/Pd surface environments, respectively, while Au$_{28}$Co$_{28}$Pd$_{29}$ represents a simplified ternary motif containing Au, Co, and Pd. The resulting cycle-level records provide the common basis for Figures \ref{fig:path_profiles}--\ref{fig:energy_shift_range}. The kMC activity scores for the comparison compositions are summarized in Table \ref{tab:control_kmc_activity} using the same $A_{\mathrm{kMC}}$ definition as in the Bayesian optimization; the score for the kMC optimum is listed in Table \ref{tab:kmc_top5}.

\begin{table}[pos=h]
    \centering
    \caption{kMC activity scores for the comparison compositions used in the path-energy analysis. The values are reported using the same $A_{\mathrm{kMC}}=\ln(\langle r\rangle)$ definition as in the kMC-based Bayesian optimization.}
    \label{tab:control_kmc_activity}
    \begin{tabular}{|l|c|}
    \hline
    Composition & $A_{\mathrm{kMC}}$ \\ \hline
    Pd$_{10}$Pt$_{58}$Rh$_{17}$ & -21.049 \\
    Pt$_{85}$ & -20.575 \\
    Au$_{43}$Pd$_{42}$ & -21.852 \\
    Co$_{43}$Pd$_{42}$ & -20.746 \\
    Au$_{28}$Co$_{28}$Pd$_{29}$ & -19.952 \\ \hline
    \end{tabular}
\end{table}

Figure \ref{fig:path_profiles} compares the ensemble path-energy profiles for these six compositions. Because the cycles contain different numbers of sampled sites, each recorded energy sequence was linearly interpolated onto a normalized path-progress coordinate from 0 (initial adsorption site) to 1 (final desorption event). This coordinate describes the order of sites sampled along a cycle and does not represent elapsed time or physical distance. In each panel, the thin lines show a randomly selected subset of individual $\mathrm{H}_2$-production cycles, and the thick line and shaded regions summarize the full cycle ensemble using the median and percentile ranges. For the kMC optimum, the pointwise median changes from $-0.065$~eV at the initial adsorption site to $0.110$~eV at the actual desorption site. Despite local fluctuations, it shows an overall shift toward weaker H binding and crosses $\Delta G_{\mathrm{H}^*}=0$ at approximately one third of the normalized path. The broad percentile ranges show that individual cycles sample diverse energy sequences around this central trend.

The static-screening optimum and pure Pt begin at stronger-binding median energies of $-0.162$ and $-0.123$~eV, respectively, and approach the thermoneutral region toward desorption, with comparatively narrow percentile ranges. Co$_{43}$Pd$_{42}$ shows a smaller change, from $-0.113$ to $-0.035$~eV, and its median remains on the stronger-binding side throughout the path. Au$_{28}$Co$_{28}$Pd$_{29}$ instead shows a broader distribution and an upward median trend from $-0.088$ to $0.081$~eV, resembling the progression observed for the kMC optimum.

For Au$_{43}$Pd$_{42}$, the median rises from $-0.196$~eV at adsorption to $0.009$~eV at desorption, while the shaded regions remain broadly extended toward negative $\Delta G_{\mathrm{H}^*}$. At the desorption endpoint, the lower quartile is $-0.164$~eV, and approximately half of the cycles terminate at negative energies.

Figure \ref{fig:ads_des_energy_distribution} compares the adsorption- and desorption-site free-energy distributions across the six compositions. For the kMC optimum, adsorption energies are concentrated on the negative-$\Delta G_{\mathrm{H}^*}$ side, whereas desorption energies are shifted toward weaker binding and span a wider range. The controls show distinct endpoint distributions: Pd$_{10}$Pt$_{58}$Rh$_{17}$ and Pt$_{85}$ have desorption energies concentrated near the thermoneutral region, Co$_{43}$Pd$_{42}$ has a negative median desorption energy, and Au$_{43}$Pd$_{42}$ retains substantial weight at negative energies.

The cycle-wise adsorption-to-desorption shifts in Figure \ref{fig:energy_shift_range} combine the two endpoint energies for each cycle. Au$_{33}$Co$_{20}$Cu$_6$Pd$_{26}$ has the broadest shift distribution in the analyzed ensembles, as measured by both the interquartile range and the 10th--90th percentile range. Its estimated density extends relatively smoothly over a wide energy interval without a dominant narrow peak, whereas several controls show stronger concentrations at particular shift values. 

\clearpage

\section{Discussion}
Static screening, which averages site-specific scores based on proximity to an optimal hydrogen adsorption energy, selected Pt-rich candidates such as Pd$_{10}$Pt$_{58}$Rh$_{17}$. The site-level analysis further showed Rh-core/Pt-shell-like ordering in the thermodynamically sampled structures, with high local activity scores at surface Pt sites. Together, these results indicate that the static objective favors Pt-rich surfaces with locally favorable hydrogen adsorption environments.

This preference for compositions containing relatively few principal elements is consistent with previous adsorption-energy-based screening studies. Pedersen et al.\cite{Pedersen2021}, for example, identified binary Ag--Pd and Ir--Pt systems in a five-element ORR search, while Xu et al.\cite{Xu2024} reported a similar tendency for adsorption-energy-based objectives.

By comparison, the kMC objective, which incorporates adsorption, diffusion, and Heyrovsky desorption rates along simulated trajectories, ranked Au$_{33}$Co$_{20}$Cu$_{6}$Pd$_{26}$ highest among the evaluated candidates. The kMC trajectory analysis supports multisite behavior in this composition. Event-frequency maps show that hydrogen adsorption occurs predominantly at Co/Pd-rich hollow sites, while Heyrovsky events occur frequently at Au-containing environments. The cycle-resolved profiles show an overall progression from stronger- to weaker-binding environments through sites with intermediate adsorption energies. The kMC optimum also exhibits the broadest, relatively smooth cycle-wise free-energy-shift distribution among the six compositions, consistent with cycles connecting a diverse range of adsorption and desorption environments. The broad adsorption-to-desorption energy shift is not, by itself, a predictor of high activity; efficient kinetics also requires that these environments be connected through accessible diffusion pathways with favorable elementary-step rates. The contrast with static screening is consistent with the inclusion of the kinetics of these modeled processes rather than site-specific adsorption energetics alone.

The division of roles among chemically distinct surface environments observed in these kMC pathways is consistent with mechanisms proposed for other compositionally complex HER catalysts. Chen et al.\cite{Chen2024} described strong-binding sites that supply adsorbed hydrogen, intermediate sites that mediate its diffusion, and weaker-binding sites that facilitate hydrogen formation and desorption on PtFeCoNiCu. He et al.\cite{He2023} studied a non-noble FeCoNiWMo high-entropy alloy with a rhomboidal structure and random atomic occupancy. They proposed that the random occupancy gives rise to a ``self-circulating adsorption-desorption structure'' and rapid internal electron transport, which together promote water-splitting activity. These studies support the broader interpretation that chemical heterogeneity can coordinate distinct reaction steps across multiple surface environments.

The present framework has several limitations, including its single-adsorbate description, the exclusion of coverage-dependent adsorbate interactions and the Tafel pathway, and the extrapolation of Pt-derived Br{\o}nsted-Evans-Polanyi parameters to multielement surface environments. Because each adsorption event begins from an unoccupied surface, the model repeatedly samples Volmer adsorption from the same type of initial state, which may underestimate the diversity of adsorption environments at finite hydrogen coverage. This protocol may also contribute to the wider desorption-energy distribution, since the desorption site is selected after a variable diffusion history. Future studies should derive element- and environment-dependent kinetic parameters from first-principles calculations and examine the effects of hydrogen coverage and competing reaction pathways. These extensions would help establish how multisite transport contributes to catalytic activity under more realistic reaction conditions.

\section{Methods}

\subsection{Machine-Learning Models for Energy Prediction}

To accelerate the large number of energy evaluations required to explore the compositional space of HEA nanoclusters, we employed a ML-based surrogate model. The model architecture, training procedures, and comprehensive performance evaluations have been reported in our previous work \cite{Otsuka_2026}. Here, we summarize the features and predictive performance most relevant to the present study.

The surrogate model is based on the Bond-type Embedded Orbital Graph Convolutional Neural Network (BE-OGCNN) architecture. A key feature of this architecture is its distance-insensitive representation, which enables efficient screening without requiring fully relaxed geometries for every candidate structure. Rather than using explicit interatomic-distance or bond-angle descriptors, BE-OGCNN represents each structure through its nearest-neighbor connectivity graph and orbital-interaction features. This representation captures chemically relevant bonding information from the elemental identities and local connectivity of the constituent atoms.

To improve data efficiency and generalization, the models were trained using a hard-parameter-sharing multi-task learning (MTL) framework. The primary task was the prediction of hydrogen adsorption energies ($\Delta E_H$) at nanocluster surface sites. Two auxiliary tasks were simultaneously learned using an extensive dataset of bulk transition-metal systems: prediction of the site-resolved d-band center ($\epsilon_d$) and total energy ($E_{\text{Total}}$). Joint learning of these physically related properties was designed to provide shared representation of the local electronic environments relevant to hydrogen adsorption and the relative energetic stability of the underlying atomic configurations.

The trained MTL models were validated using 85-atom nanoclusters with ternary compositions that were excluded from the training data. The training data contained 19- and 44-atom clusters and publicly available slab datasets. As detailed in the Supporting Information, the hydrogen-adsorption-energy model achieved a mean absolute error (MAE) of 0.105 eV for the 85-atom test clusters (Figure S1), while the total energy model achieved a root-mean-square error (RMSE) of 48.173 meV/atom (Figure S2). These models were subsequently used for the energy evaluations required in the composition-optimization framework.

\subsection{Structural Model and Thermodynamic Sampling of Atomic Configurations}

As the target for composition optimization, we adopted a five-layer, face-centered cubic (FCC) octahedral cluster consisting of 85 atoms and having a geometric diameter of approximately 1.7 nm. The cluster is enclosed by close-packed (111) facets, which are relevant to catalytic applications\cite{Wang2013}. Although truncated-octahedral morphologies are generally favored for larger FCC nanoparticles, density functional theory (DFT) calculations for single-element clusters showed that regular and truncated octahedral structures are energetically comparable at the four-layer (44-atom) and five-layer (85-atom) sizes, with no consistent preference for either morphology across the elements considered (Supporting Information, Section S4). The regular octahedral cluster therefore provides a reasonable structural model for the present study.

This cluster size was selected based on both computational feasibility and experimental relevance. Feng et al.\cite{Feng2021} synthesized HEA nanoparticles with an average diameter of 1.68 nm, corresponding approximately to an edge length of four to five atomic layers, and reported high HER activity and stable performance over 10,000 cycles. These observations support the use of the 85-atom cluster as a representative model of ultrasmall HEA nanoparticles.

Nine transition-metal elements were selected as candidate constituents: Ag, Au, Co, Cu, Ni, Ir, Pd, Pt, and Rh. This nine-element chemical space was explored to identify active HER compositions while accounting for interactions among the constituent elements.

In HEA nanoparticles of this size, differences in elemental interactions can give rise to surface segregation, chemical ordering, and elemental clustering. These effects can alter the surface composition and local atomic environments relevant to catalytic activity. We therefore sampled thermodynamically accessible atomic configurations rather than assuming a random solid solution for each composition.

MC simulations were performed to sample atomic configurations of the 85-atom cluster at each fixed composition. At each MC step, two atoms of different species were selected randomly, and their positions were exchanged. The trial configuration was accepted according to the Metropolis criterion:
\begin{equation}
P = \min\left(1, \exp\left(-\frac{\Delta E}{k_B T}\right)\right)
\end{equation}
where 
\begin{equation}
\Delta E = E_{\text{trial}} - E_{\text{current}},
\end{equation}
$E_{\text{trial}}$ and $E_{\text{current}}$ are the BE-OGCNN-predicted total energies of the trial and current configurations, respectively, $k_B$ is the Boltzmann constant, and $T$ is the sampling temperature.

Based on the previous study by Huang et al.\cite{Huang2024}, the sampling temperature was set to 900 K. This temperature is representative of annealing conditions used in the synthesis of HEA nanoparticles and was selected to sample elemental distributions relevant to thermally processed nanoparticles. Sampling at 900 K also facilitates configurational exploration by increasing the acceptance probability of energetically unfavorable trial swaps compared with sampling at room temperature.

For each composition, 6,000 MC steps were performed to balance computational cost and configurational sampling. The first 5,000 steps were excluded as an equilibration period, and the configurations generated during the final 1,000 steps were used for the subsequent property evaluations.

\subsection{Static HER Activity Metric}
Using the thermodynamically accessible atomic configurations generated by the MC sampling described above, we defined a static activity metric for composition screening and examined its consistency with the established HER activity trend of pure metals.

Because nanocluster surfaces contain adsorption sites with diverse local elemental and geometric environments, including terraces, edges, and corners, their activity cannot be adequately represented by a single atomic configuration or adsorption site. We therefore evaluated each composition by averaging site-specific activity scores over the sampled atomic configurations and adsorption sites, as described below.

\begin{enumerate}
    \item \textbf{Configurational Sampling:} For each composition, 1,000 atomic configurations were sampled from the equilibrated final portion of the MC trajectory at 900 K and used for activity evaluation.
    \item \textbf{Adsorption Energy Calculation:} For each sampled atomic configuration, all identified surface adsorption sites, including on-top, hollow, and bridge sites, were considered, and the hydrogen adsorption energy $\Delta E_{\mathrm{H},i}$ at each site $i$ was predicted using the BE-OGCNN model.
    \item \textbf{Site-Specific Activity Calculation:} Based on the Sabatier principle, optimal HER activity is expected when hydrogen binds neither too strongly nor too weakly. Following the activity model proposed by Pedersen et al.\cite{Pedersen2021}, the local activity score $a_i$ at each adsorption site $i$ was defined as
    \begin{equation}
        a_i = \exp\left(-\frac{|\Delta E_{\mathrm{H},i} - \Delta E_{\mathrm{opt}}|}{k_B T}\right)
    \end{equation}
    where $\Delta E_{\mathrm{opt}}$ is the optimal adsorption energy for HER activity, $k_B$ is the Boltzmann constant, and $T$ is the reaction temperature. In this study, $\Delta E_{\mathrm{opt}}$ and $T$ were set to $-0.27$ eV and 298 K, respectively. With the $+0.24$~eV free-energy correction used below, this adsorption-energy optimum corresponds to $\Delta G_{\mathrm{H}^*}\approx-0.03$~eV. The value of $\Delta E_{\mathrm{opt}}$ was determined using the thermodynamic corrections described in Ref.\cite{Tran2018}.
    \item \textbf{Overall Activity Calculation:} The overall static activity metric $A$ for each composition was defined as the average of the local activity scores over all adsorption sites and sampled atomic configurations:
    \begin{equation}
        A = \frac{1}{N_{\mathrm{total}}} \sum_{i=1}^{N_{\mathrm{total}}} a_i
    \end{equation}
    where $N_{\mathrm{total}}$ is the total number of adsorption sites evaluated across all sampled atomic configurations.
\end{enumerate}

To assess whether the activity metric $A$ reproduces established elemental trends, we evaluated 85-atom clusters of each of the nine constituent metals. As detailed in Supporting Information Section S5, the predicted activity trend was in qualitative agreement with the experimentally established HER volcano relationship\cite{Nerskov2005}. This agreement supports the use of $A$ as a relative screening metric for comparing the modeled nanocluster compositions.

\subsection{Optimization Algorithms}
To efficiently explore the large compositional space, we compared two optimization approaches implemented using PHYSBO and Optuna.

\textbf{PHYSBO (candidate-based Bayesian optimization):} PHYSBO\cite{Motoyama2022} is a Python library for Bayesian optimization that sequentially selects candidates from a predefined pool according to an acquisition function. We constructed a pool of 500,000 unique compositions, each containing exactly 85 atoms, using a randomized procedure. First, the inclusion of each of the nine elements was determined using a binary flag. For each included element, an integer weight between 1 and 85 was then sampled from a uniform distribution. The sampled weights were normalized and converted to integer atom counts whose sum was constrained to 85, with the residual arising from rounding assigned to one of the included elements. Pure-metal compositions and duplicate candidates were removed, ensuring that each candidate contained at least two elements. This randomized procedure generated candidates with varying numbers and proportions of constituent elements without restricting the compositions to a predefined coarse grid, such as 10 at.\% intervals.

\textbf{Optuna (Tree-structured Parzen Estimator, TPE):} Optuna\cite{Akiba2019} is an optimization framework that implements the TPE algorithm and can handle mixed and conditional search spaces. For each of the nine elements, we introduced (i) a binary categorical variable indicating whether the element was included and (ii) a continuous variable representing its relative amount. Optuna therefore explored a total of 18 variables, comprising nine binary inclusion variables and nine continuous composition variables. The continuous amounts of the included elements were normalized and converted to integer atom counts under the constraint that the total number of atoms was 85.

\subsection{Dynamic Activity Evaluation via Kinetic Monte Carlo}
\label{sec:kmc_method}

To evaluate the catalytic activity of HEA catalysts under dynamic conditions, including the surface diffusion of adsorbates, a kMC simulation model was developed based on the framework reported by Gu et al.\cite{Gu2021}.

\subsubsection{Simulation Model and Elementary Processes}

Because the ML model was developed to predict the adsorption energies of a single hydrogen adsorbate on the nanocluster surface, coverage-dependent adsorbate--adsorbate interactions cannot be directly evaluated within this framework. Therefore, assuming a low-overpotential and low-hydrogen-coverage regime, the simulation tracks the dynamic behavior of a single hydrogen atom through three elementary processes and explicitly excludes the bimolecular Tafel recombination mechanism ($2\mathrm{H}_{\mathrm{ad}} \rightarrow \mathrm{H}_2$):

\begin{enumerate}
    \item \textbf{Adsorption (Volmer process):} A proton and an electron from the environment react to form an adsorbed hydrogen atom: $\mathrm{H}^+ + \mathrm{e}^- \rightarrow \mathrm{H}_{\mathrm{ad}}$. Adsorption at each available surface site is treated as a possible kMC event.
    \item \textbf{Surface diffusion:} The adsorbed hydrogen atom hops to an adjacent adsorption site on the surface.
    \item \textbf{Desorption (Heyrovsky process):} The adsorbed hydrogen atom reacts with another proton and an electron to desorb as a hydrogen molecule: $\mathrm{H}_{\mathrm{ad}} + \mathrm{H}^+ + \mathrm{e}^- \rightarrow \mathrm{H}_2$.
\end{enumerate}

The kMC simulation progresses by probabilistically selecting an event from those available in the current state according to their respective rate constants. Volmer adsorption events are considered when the surface is unoccupied, whereas diffusion and Heyrovsky desorption events compete after hydrogen adsorption. In this study, each kMC trajectory consisted of 40,000 event transitions.

\subsubsection{Reaction Rate Constants and Activity Metric}

The rate constants for the elementary events in the kMC simulations were calculated using the following transition-state-theory expression:
\begin{equation}
    k = \frac{k_B T}{h} \exp\left(-\frac{\Delta G_{\mathrm{act}}}{k_B T}\right),
\end{equation}
where $h$ is Planck's constant and $\Delta G_{\mathrm{act}}$ is the activation free energy. 

The activation free energy for each process was estimated using Br{\o}nsted-Evans-Polanyi (BEP) relations\cite{Bligaard2004, Michaelides2003} based on the local hydrogen adsorption free energies:
\begin{align}
    \Delta G_{\mathrm{act,Vol}} &= \Delta G_{0,\mathrm{Vol}}^\dagger + \beta_{\mathrm{Vol}} \Delta G_{\mathrm{H}^*} , \\
    \Delta G_{\mathrm{act,diff}} &= \Delta G_{0,\mathrm{diff}}^\dagger + \beta_{\mathrm{diff,1}} \Delta G_{\mathrm{H}^*,\mathrm{final}} - \beta_{\mathrm{diff,2}} \Delta G_{\mathrm{H}^*,\mathrm{initial}} , \\
    \Delta G_{\mathrm{act,Hey}} &= \Delta G_{0,\mathrm{Hey}}^\dagger + \beta_{\mathrm{Hey}} (-\Delta G_{\mathrm{H}^*}) .
\end{align}

Here, $\Delta G_0^\dagger$ represents the intrinsic activation barrier at the equilibrium potential, and $\beta$ is the corresponding BEP coefficient. The hydrogen adsorption free energy at a given site, $\Delta G_{\mathrm{H}^*}$, was calculated from the ML-predicted adsorption energy, $\Delta E_{\mathrm{H}}$, as $\Delta G_{\mathrm{H}^*} = \Delta E_{\mathrm{H}} + 0.24$ eV, where the correction accounts for zero-point-energy and entropic contributions, following N{\o}rskov et al.\cite{Nerskov2005}.

The BEP parameters, $\Delta G_0^\dagger$ and $\beta$, were adopted from the computational study by Gu et al.\cite{Gu2021}. Although these BEP parameters were originally derived for pure Pt surfaces, we extrapolated them to estimate transition barriers and the corresponding transition rates across the chemically diverse local environments of multicomponent alloy surfaces. This application extends beyond the chemical domain for which the BEP parameters were originally established and therefore represents an approximation of the present methodology. Pure-metal nanoclusters were used to check the qualitative activity trends of the resulting model, as described below and in the Supporting Information (Section S8).

In such surface reactions, a severe time-scale disparity can arise because diffusion barriers are lower than desorption barriers. To reduce the computational cost of frequent diffusion events, a dynamic rate constant rescaling method\cite{Nunez2017} was applied. The detailed methodology and scaling algorithm are provided in the Supporting Information (Section S7). The rescaling uses a target diffusion-to-desorption event ratio of $N_{\mathrm{min}} = 30$, with the aim of retaining fast diffusion relative to desorption. This target is an algorithmic control parameter rather than a guaranteed minimum for every cycle. Provided that rescaled diffusion remains much faster than desorption ($k'_{\mathrm{diff}} \gg k_{\mathrm{des}}$), the rescaling is expected to have only a limited effect on the computed H$_2$ production rate.
Before applying this kMC framework to multicomponent compositions, we evaluated 85-atom pure-metal nanoclusters. As discussed in the Supporting Information (Section S8), the resulting activity trend was qualitatively consistent with the experimentally established HER volcano relationship across transition metals\cite{Nerskov2005}. This comparison supports the use of the model for exploratory screening, but does not independently validate either the transferability of the Pt-derived BEP parameters or the accuracy of the rate-rescaling approximation on multicomponent surfaces.

For each sampled atomic configuration $j$, the $\mathrm{H}_2$ production rate was calculated as
\begin{equation}
    r_j = \frac{N_{\mathrm{H}_2,j}}{t_{\mathrm{total},j}}.
\end{equation}
Here, $N_{\mathrm{H}_2,j}$ is the number of H$_2$ molecules produced for configuration $j$, and $t_{\mathrm{total},j}$ is the total elapsed kMC simulation time for that configuration.

To account for the diversity of atomic configurations in HEA nanoparticles, the composition-level production rate was calculated as the arithmetic mean of $r_j$ over the 1,000 configurations sampled by MC at 900 K:
\begin{equation}
    \langle r \rangle = \frac{1}{N_{\mathrm{conf}}}
    \sum_{j=1}^{N_{\mathrm{conf}}} r_j.
\end{equation}
Because the rates span many orders of magnitude, the dynamic activity score used for composition optimization was defined as the natural logarithm of this configurationally averaged rate:
\begin{equation}
    A_{\mathrm{kMC}} =
    \ln\left(\langle r \rangle\right)
    = \ln\left[
    \frac{1}{N_{\mathrm{conf}}}
    \sum_{j=1}^{N_{\mathrm{conf}}}
    r_j
    \right],
\end{equation}
where $N_{\mathrm{conf}}=1000$. Repeated configurations were evaluated once and weighted by their occurrence counts, which is algebraically equivalent to calculating the arithmetic mean over the sampled configuration sequence.

For the mechanistic path-energy analysis, each completed $\mathrm{H}_2$-production cycle was decomposed into the sequence of adsorption sites visited by the hydrogen atom from the initial Volmer adsorption event to the final Heyrovsky desorption event. Consecutive revisits were omitted, and the sites visited for the first time were retained in chronological order. The actual desorption site was then placed at the end of the sequence; when that site had been visited earlier, it therefore appeared once in its first-visit position and again as the terminal reaction site. The resulting sequence was assigned equally spaced coordinates from 0 at the initial adsorption site to 1 at the final desorption event. The site energies were linearly interpolated onto 101 equally spaced points along this normalized coordinate, allowing ensemble profiles with different numbers of sites to be compared on a common axis. The coordinate describes relative progress through the site sequence and does not encode elapsed time, hop rate, or physical travel distance. Pointwise ensemble medians and 10th--90th and 25th--75th percentile ranges were calculated over all analyzed cycles. For each cycle, the adsorption-to-desorption free-energy shift was calculated as $\Delta G_{\mathrm{des}}-\Delta G_{\mathrm{ads}}$, using the free energies at the initial adsorption and final desorption sites, respectively. Adsorption energies, desorption energies, and cycle-wise shifts were visualized using violin plots, with the median and interquartile range shown explicitly.

\subsection{Use of Generative AI}

During the preparation and revision of this manuscript, the authors used ChatGPT and Codex (OpenAI) to assist with language editing, manuscript organization, code review, and the preparation of analysis and visualization scripts. All AI-assisted outputs were critically reviewed and, where necessary, modified by the authors. The authors verified the analyses, numerical results, references, and final manuscript and take full responsibility for the published content.

\section*{Acknowledgements}
This work was supported by the Japan Science and Technology Agency (JST) as part of SICORP (Grant Number JPMJSC21E2), and by the Japan Society for the Promotion of Science (JSPS) KAKENHI (Grant Number 24K01284). Part of the calculations were performed using the supercomputers
at the Information Technology Center and the Institute for Solid State Physics (ISSP), the University of Tokyo.

\section*{Competing Interests}
The authors report there are no competing interests to declare.

\section*{Author contributions}
Koki Otsuka: Conceptualization, Methodology, Software, Investigation, Formal analysis, Writing - original draft. Koji Shimizu: Investigation, Data curation. Anh Khoa Augustin Lu: Writing - review and editing. Satoshi Watanabe: Writing - review and editing, Supervision.

\section*{Data availability}
The code, including the custom kinetic Monte Carlo and thermodynamic sampling scripts, and the optimization frameworks used in this study are openly available on GitHub at \url{https://github.com/modelna/NanoAlloy-Opt.git}.

\bibliographystyle{elsarticle-num}
\bibliography{references1}

\end{document}

\let\WriteBookmarks\relax
\def\floatpagepagefraction{1}
\def\textpagefraction{.001}

\shorttitle{Supporting Information}
\shortauthors{Otsuka et al.}

\title[mode=title]{Supporting Information: Dynamic Kinetic Evaluation Favors Compositionally Diverse Multicomponent Alloy Nanoparticles for Hydrogen Evolution}

\author[1]{Koki Otsuka}
\author[1,2]{Anh Khoa Augustin Lu}
\author[3]{Koji Shimizu}
\author[1]{Satoshi Watanabe}
\cormark[1]
\ead{watanabe@cello.t.u-tokyo.ac.jp}

\affiliation[1]{organization={Department of Materials Engineering, The University of Tokyo},
            postcode={113-8656},
            city={Tokyo},
            country={Japan}}

\affiliation[2]{organization={Research Center for Materials Nanoarchitectonics, National Institute for Materials Science},
            postcode={305-0044},
            city={Tsukuba},
            country={Japan}}

\affiliation[3]{organization={Materials DX Research Center, National Institute of Advanced Industrial Science and Technology},
            postcode={305-8568},
            city={Tsukuba},
            country={Japan}}

\cortext[1]{Corresponding author}

\maketitle

\section*{S1. Machine Learning Model Overview}
The predictive model developed in this study is an enhanced distance-insensitive graph neural network named the Bond-type Embedded Orbital Graph Convolutional Neural Network (BE-OGCNN).  Built upon the BE-CGCNN framework \cite{Bang2023}, this model is designed to rely solely on the connectivity graph of nearest-neighbor atoms, bypassing the need for precise geometric inputs such as bond lengths and angles. To maximize expressivity without geometric dependency, the BE-OGCNN integrates orbital interaction features. 

Furthermore, the model employs a hard parameter sharing Multi-Task Learning (MTL) framework to improve data efficiency and generalization. The primary task is the prediction of hydrogen adsorption energy, while the d-band center and total structural energy serve as auxiliary tasks. 

\clearpage
\section*{S2. Datasets}
The model was trained and evaluated using a combination of three distinct datasets:
\begin{itemize}
    \item \textbf{In-House Cluster Dataset:} Constructed using Density Functional Theory (DFT) calculations, this dataset contains hydrogen adsorption energies on composed octahedral cluster models.  It includes nine transition metal elements (Ag, Au, Co, Cu, Ir, Ni, Pd, Pt, Rh) across various cluster sizes: 19-atom (3,820 data points), 44-atom (4,254 data points), and 85-atom (115 data points) models. The 85-atom data were reserved exclusively for testing and were not used for model training.
    \item \textbf{OpenCatalyst 20 (OC20):} A subset of 1,316 data points containing hydrogen adsorption energies and total energies for various transition metal surfaces was extracted for the primary and auxiliary tasks \cite{Chanussot2021}.
    \item \textbf{Alexandria Dataset:} Used for the auxiliary tasks, this dataset provided total energy data for bulk crystal structures of pure transition metals and their alloys \cite{Cavignac2026Alexandria}. The corresponding d-band center values were obtained by parsing VASP output files from the NOMAD repository.
\end{itemize}

\clearpage
\section*{S3. Performance Validation on 85-Atom Five-Layer Octahedral Clusters}

To evaluate the extrapolation capability and transferability of our models to more realistic catalyst sizes, performance validation was conducted on 85-atom (five-layer) octahedral clusters. It is important to emphasize that this 85-atom cluster size represents a structural domain entirely absent from the training dataset.

\subsubsection*{Hydrogen Adsorption Energy Prediction}
\begin{figure}[pos=h]
    \centering
    \includegraphics[width=0.55\textwidth]{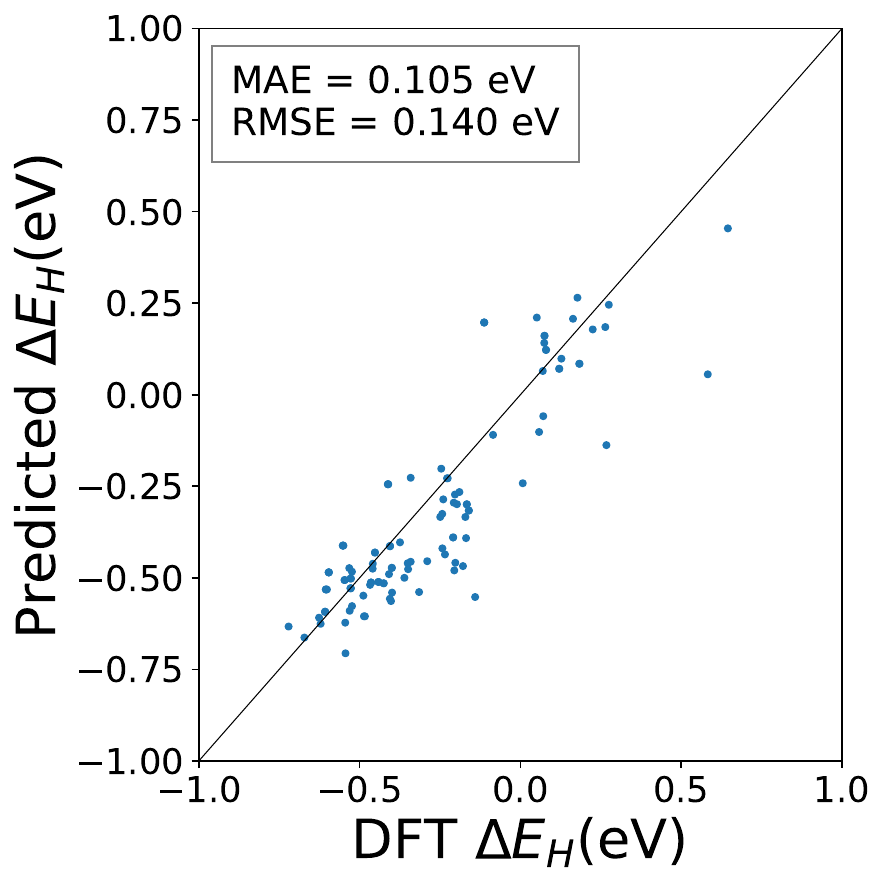} 
    \caption{Parity plot comparing the DFT-calculated and ML-predicted hydrogen adsorption energies ($\Delta E_{\text{H}}$) for the 85-atom octahedral clusters. The multi-task learning (MTL) model demonstrates robust extrapolation performance on structures larger than those included in the training set.}
    \label{fig:S_adsorption}
\end{figure}

For the prediction of hydrogen adsorption energy ($\Delta E_{\text{H}}$), the model was trained strictly on smaller nanoclusters (19-atom and 44-atom), surface slab models, and bulk crystal data. When evaluated on the unseen 85-atom clusters, the results demonstrated that the multi-task learning (MTL) approach significantly enhances predictive reliability for unknown, larger structures. Specifically, the MTL model achieved a Mean Absolute Error (MAE) of 0.105 eV, showing a distinct improvement over the single-task model, which yielded an MAE of 0.122 eV. Figure \ref{fig:S_adsorption} illustrates the correlation between the DFT-calculated and ML-predicted hydrogen adsorption energies.

\subsubsection*{Total Energy Prediction for Compositional Optimization}
In addition to adsorption energy, accurately predicting the total structural energy ($E_{\text{Total}}$) is crucial for evaluating structural stability and performing Monte Carlo sampling during composition optimization. To this end, a total energy prediction model was also developed using the BE-OGCNN architecture. 

To ensure the diversity of local atomic environments and improve size-extrapolation capability, the training dataset for this model consisted of 4,000 structures, evenly sampled from four different sources (1,000 structures each): 19-atom octahedral clusters, 44-atom octahedral clusters, surface slab models, and bulk crystals. 

\begin{figure}[pos=h]
    \centering
    \includegraphics[width=0.55\textwidth]{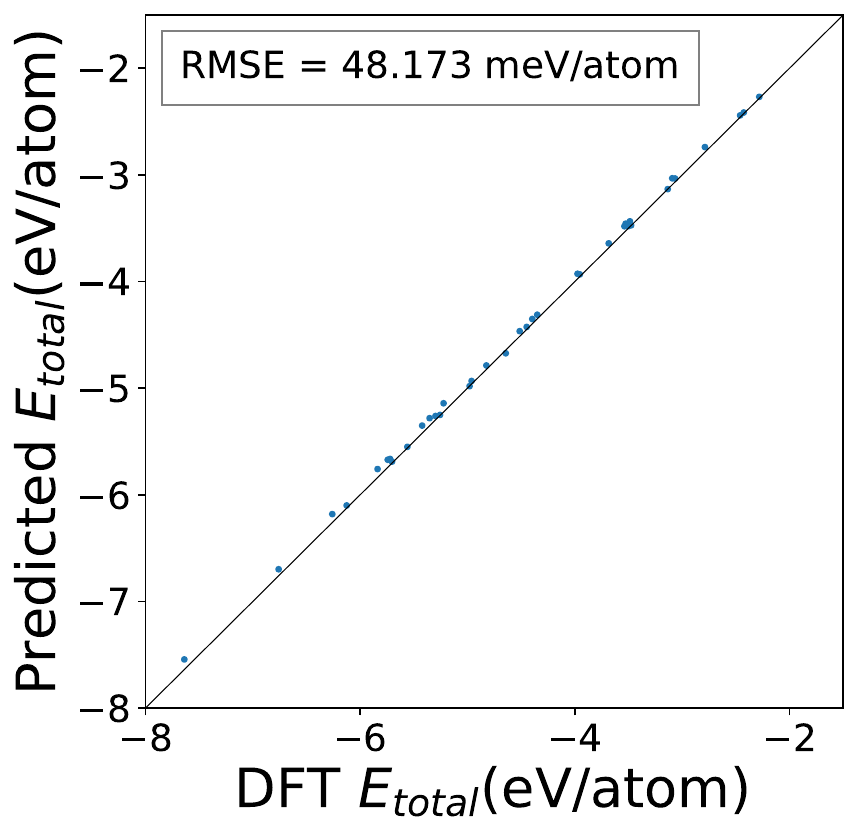} 
    \caption{Parity plot comparing the DFT-calculated and ML-predicted total energies per atom for the 85-atom octahedral test clusters, which are larger than the clusters used for training. The RMSE is 48.173 meV/atom.}
    \label{fig:S_total_energy}
\end{figure}

When this model was tested on the unseen 85-atom clusters, the predicted total energies per atom achieved a root-mean-square error (RMSE) of 48.173 meV/atom relative to DFT (Figure \ref{fig:S_total_energy}). This test characterizes prediction errors for the larger cluster size. It does not directly quantify errors in the energy differences between fixed-composition configurations that determine MC swap acceptance probabilities.

\clearpage
\section*{S4. Stability Comparison: Octahedral vs.\ Truncated Octahedral Clusters}

A commonly raised question regarding the choice of cluster geometry is whether the regular octahedral structure is energetically favorable compared to the truncated octahedral structure at the sizes relevant to this study. To address this point, density functional theory (DFT) calculations were performed to compare the per-atom energies of regular octahedral and truncated octahedral cluster models for each of the nine single-element candidate metals, at both the four-layer and five-layer sizes.

The DFT calculations were performed using the Vienna Ab initio Simulation Package (VASP)\cite{Kresse1993}, employing the Perdew-Burke-Ernzerhof (PBE)\cite{Perdew1996} exchange-correlation functional and the projector-augmented-wave (PAW)\cite{Blochl1994} method. A plane-wave energy cut-off of 500 eV was used. Brillouin-zone integrations were performed using Monkhorst-Pack k-point sampling\cite{Monkhorst1976}, with $3\times3\times3$ grids for the four-layer (44-atom) models and $1\times1\times1$ grids for the five-layer (85-atom) models. A vacuum spacing of at least 15~\AA~ was introduced to avoid interactions between periodic images. The electronic convergence criterion was set to $10^{-4}$ eV, and the atomic configurations were relaxed until the atomic forces were below 0.01 eV/\AA.

\begin{figure}[pos=h]
    \centering
    \includegraphics[width=0.65\textwidth]{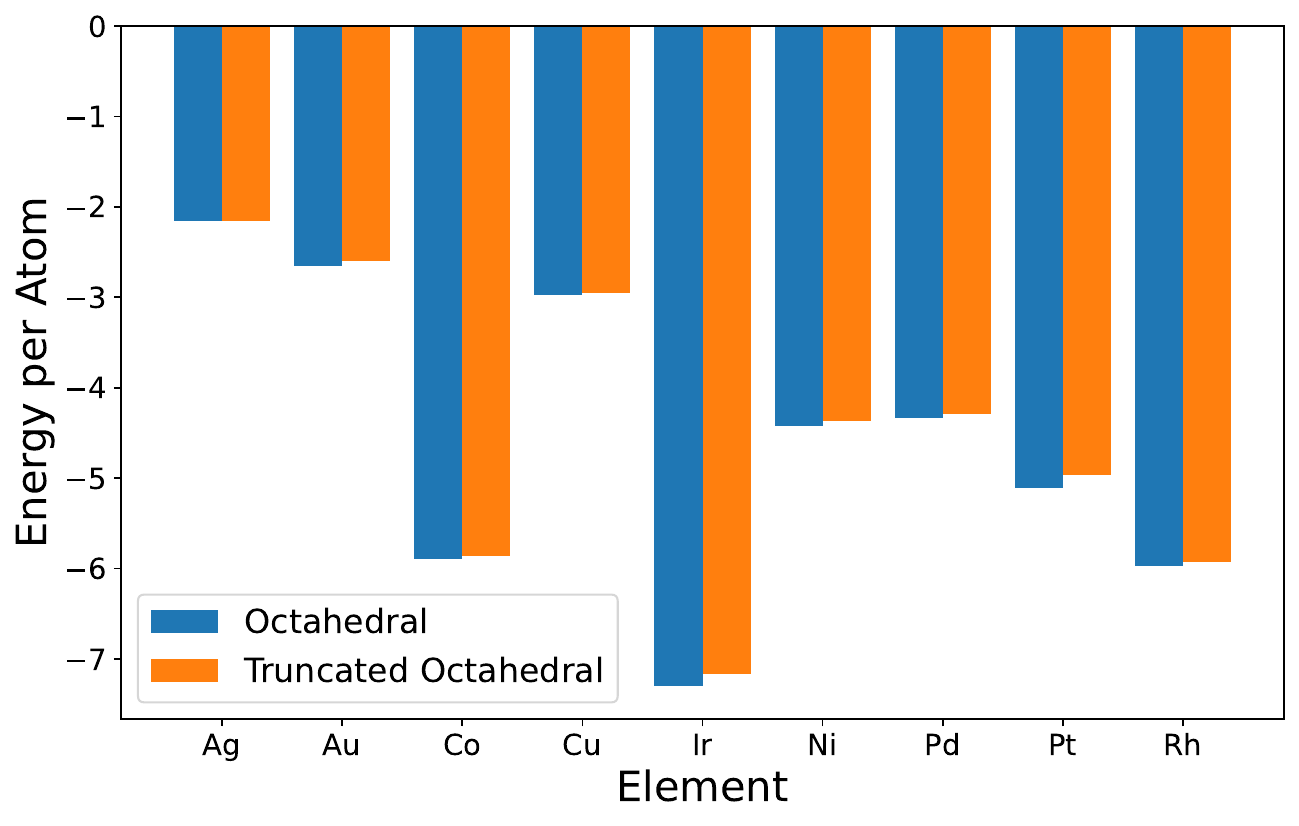}
    \caption{DFT-calculated per-atom energies for regular octahedral (blue) and truncated octahedral (orange) clusters at the four-layer size for each of the nine single-element candidate metals. Lower energy indicates greater structural stability.}
    \label{fig:S_stability_4layer}
\end{figure}

\begin{figure}[pos=h]
    \centering
    \includegraphics[width=0.65\textwidth]{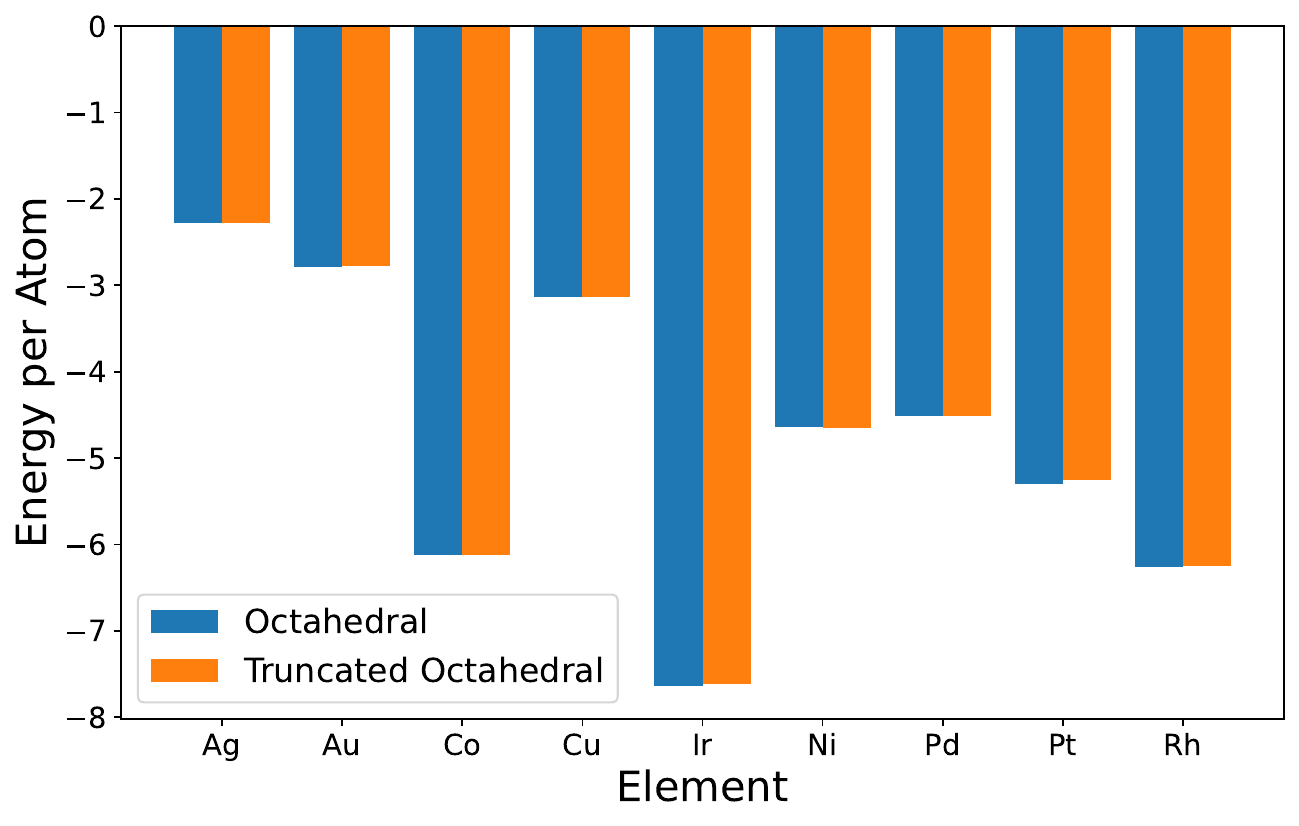}
    \caption{DFT-calculated per-atom energies for regular octahedral (blue) and truncated octahedral (orange) clusters at the five-layer size for each of the nine single-element candidate metals. Lower energy indicates greater structural stability.}
    \label{fig:S_stability_5layer}
\end{figure}

Figures \ref{fig:S_stability_4layer} and \ref{fig:S_stability_5layer} show the DFT-calculated per-atom energies for the four-layer and five-layer cases, respectively.

As shown in the figures, the per-atom energies of the regular octahedral and truncated octahedral clusters are remarkably close for most elements, and no decisive energetic preference for either shape is apparent across all nine metals. It is well established that for large FCC metal nanoparticles the truncated octahedron is the thermodynamically preferred equilibrium shape, because truncation reduces the contribution of energetically costly edge and vertex atoms \cite{Barnard2005}. However, at small sizes (sub-2 nm regime), the energy differences between competing shapes become progressively smaller and the relative stability can vary with element and exact atom count. The results here are consistent with this size-dependent picture: the two geometries are energetically comparable at the four-layer and five-layer sizes studied, and neither structure is unambiguously more stable for all elements. Given that the energy differences are marginal and that the experimental synthesis by Feng et al. \cite{Feng2021} produced clusters with an average diameter of 1.68 nm consistent with the present model, the regular octahedral cluster is a physically reasonable structural model for this study.

\clearpage
\section*{S5. Validation of the Activity Metric using Pure Metal Clusters}

To assess the elemental trend of the static activity metric $A$ defined in the main text, we evaluated 85-atom pure-metal clusters for each of the nine candidate transition metals (Ag, Au, Co, Cu, Ir, Ni, Pd, Pt, and Rh). The score was averaged over the available adsorption sites, including their different local geometric environments. 

\begin{figure}[pos=h]
    \centering
    \includegraphics[width=0.55\textwidth]{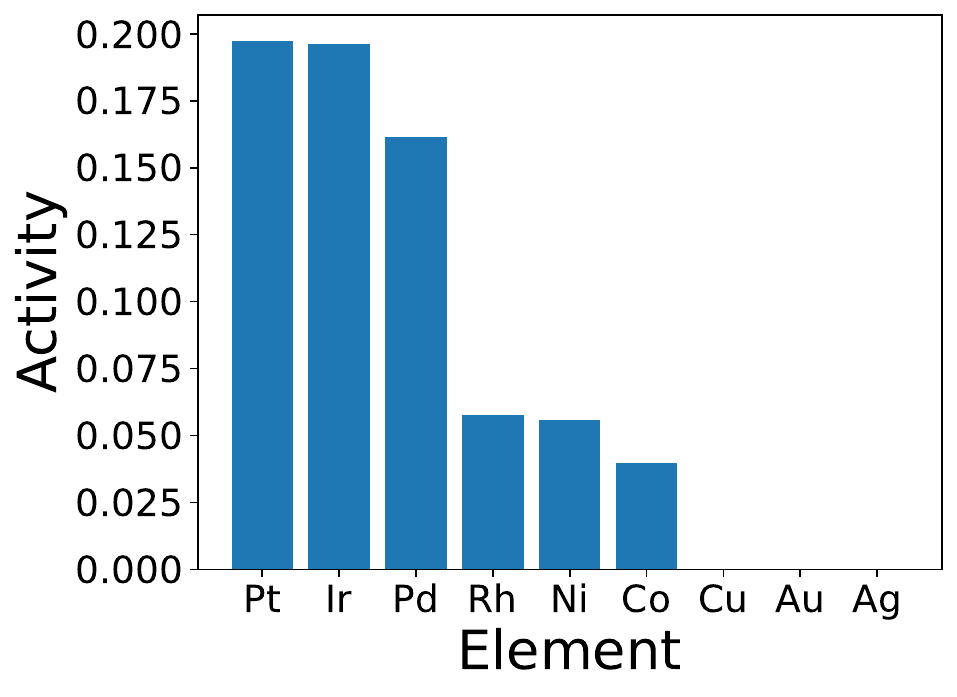} 
    \caption{Calculated static HER activity scores ($A$) for 85-atom pure transition-metal clusters. The qualitative activity trend is compared with the established HER volcano relationship as a check of the screening metric.}
    \label{fig:S_volcano}
\end{figure}

Figure \ref{fig:S_volcano} illustrates the calculated static activity metric $A$ for each pure metal cluster.  The evaluation results demonstrated that Pt exhibited the highest activity, followed by Ir and Pd, whereas elements like Au, Ag, and Cu showed significantly lower activities. This sequence is in excellent qualitative agreement with the experimentally established volcano plot trend for the hydrogen evolution reaction (HER). The agreement supports the use of this activity evaluation as a relative screening metric for comparing the modeled nanocluster compositions, while its transferability to more complex multicomponent surfaces remains an approximation.

\clearpage
\section*{S6. Parameters for Brønsted-Evans-Polanyi (BEP) Relations}

In the main text, the activation free energies ($\Delta G_{act}$) for the elementary processes in the kinetic Monte Carlo (kMC) simulations are calculated based on Brønsted-Evans-Polanyi (BEP) relations. The intrinsic activation barriers ($\Delta G_{0}^{\dagger}$) and the symmetry factors ($\beta$) for these linear scaling relations were adopted from the computational work by Gu et al. \cite{Gu2021}

The specific parameter values used in our simulations for each elementary step are as follows:
\begin{itemize}
    \item \textbf{Adsorption (Volmer process):} The intrinsic activation barrier and symmetry factor are set to $\Delta G_{0,Vol}^{\dagger} = 1.173$ eV and $\beta_{Vol} = 0.881$, respectively.
    \item \textbf{Surface Diffusion:} The intrinsic activation barrier for diffusion is $\Delta G_{0,diff}^{\dagger} = 0.184$ eV. Because the activation energy for surface diffusion depends on the hydrogen adsorption free energies of both the initial and final sites, two distinct symmetry factors are applied to govern the energy transitions: $\beta_{diff,1} = 0.435$ and $\beta_{diff,2} = 0.565$. 
    \item \textbf{Desorption (Heyrovsky process):} The intrinsic activation barrier and symmetry factor are set to $\Delta G_{0,Hey}^{\dagger} = 1.360$ eV and $\beta_{Hey} = 0.511$, respectively. These values appropriately reflect the relatively high activation barrier associated with the release of a hydrogen molecule compared to surface diffusion.
\end{itemize}
These parameterized BEP relations provide an efficient approximate means of estimating transition rates across the diverse local environments considered here.

\clearpage
\section*{S7. Rescaling of Reaction Rate Constants for Time-Scale Disparity}

In the present HER model, the intrinsic barrier parameters for diffusion and Heyrovsky desorption are 0.184 and 1.360 eV, respectively. The actual activation barriers also depend on the local adsorption free energies through the BEP relations in Section S6; these intrinsic values are not fixed barriers for every site. When diffusion is much faster than desorption, most kMC event selections are diffusion hops, making the observation of completed reaction cycles computationally expensive.

To resolve this issue, a rate constant rescaling method was applied \cite{Nunez2017}. This technique artificially divides the rate constant of the fast process (diffusion) by a dynamic scaling factor $\alpha$, thereby narrowing the kinetic gap with the slow process (desorption) and improving computational efficiency while retaining the intended fast-diffusion approximation.

The scaling factor $\alpha$ is dynamically updated every 2,000 kMC steps according to the following equations:
\begin{align}
    N &= \frac{0.5 \times (F_{fwd} + F_{bwd})}{F_{slow}} \\
    \alpha &= \frac{N}{N_{min}}\\
    k'_{diff} &= \frac{k_{diff}}{\alpha} \quad (\text{applied only when } \alpha > 1) \\
\end{align}
where $F_{fwd}$ and $F_{bwd}$ represent the total numbers of forward and backward hops observed for a given diffusion pair (i.e., transitions between an occupied site and one of its neighboring vacant sites) in the preceding 2,000 kMC steps, and $F_{slow}$ is the number of desorption events during the same interval. The numerator is the arithmetic mean of the forward and backward hop counts; it does not measure statistically independent excursions. The rescaling is applied only when $\alpha > 1$, i.e., when the event-count ratio $N$ exceeds the target value $N_{min}$. No additional reduction is applied when $\alpha \le 1$. The target was set to $N_{min} = 30$ to retain diffusion sampling before desorption; this setting does not guarantee a minimum number of hops in each cycle or establish convergence of surface exploration.

\clearpage
\section*{S8. Kinetic Monte Carlo Evaluation of Pure Metal Clusters}

Before applying the kinetic Monte Carlo (kMC) framework to multicomponent alloys, its qualitative activity trends were examined using 85-atom pure-metal nanoclusters. The static evaluation in Section S5 averages site-specific activity scores, rather than adsorption energies themselves. In contrast, the kMC approach explicitly models surface diffusion and the competition between elementary events, with transition barriers estimated using Brønsted-Evans-Polanyi (BEP) relations.

\begin{figure}[pos=h]
    \centering
    \includegraphics[width=0.55\textwidth]{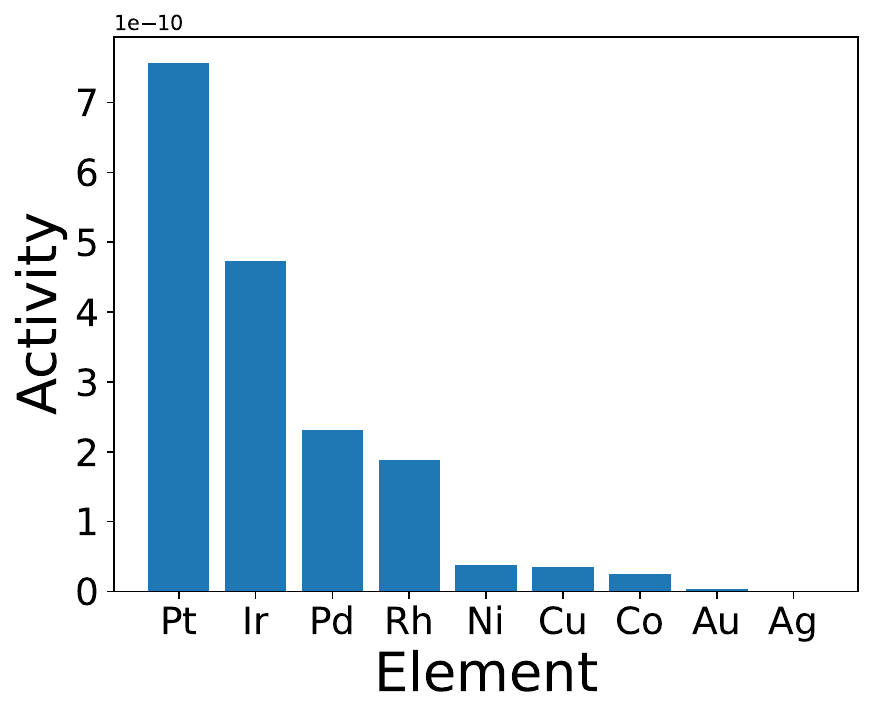} 
    \caption{H$_2$ production rates, $r=N_{\mathrm{H}_2}/t_{\mathrm{total}}$, for 85-atom pure transition-metal clusters obtained from kMC simulations. The figure is used to compare the qualitative activity trend among the pure metals; the plotted rate is not on the same numerical scale as the logarithmic $A_{\mathrm{kMC}}$ values reported in the main text.}
    \label{fig:S_kmc_pure}
\end{figure}

For the pure-metal benchmark, Figure \ref{fig:S_kmc_pure} reports the H$_2$ production rate rather than the logarithmic activity score used in the main text. The plotted quantity was calculated as
\begin{equation}
    r = \frac{N_{\mathrm{H}_2}}{t_{\mathrm{total}}}
\end{equation}
where $N_{\mathrm{H}_2}$ is the number of H$_2$ molecules produced and $t_{\mathrm{total}}$ is the total elapsed kMC simulation time. As in the main-text calculations, each kMC trajectory comprised 40,000 event transitions. Figure \ref{fig:S_kmc_pure} is used to compare the qualitative activity trend among the pure metals; its rate values are not on the same numerical scale as the logarithmic $A_{\mathrm{kMC}}$ values reported in the main text.

Similar to the static evaluation results, the pure-metal kMC benchmark placed Pt in the highest-activity group, together with Ir and Pd. Conversely, elements with very weak or strong hydrogen-binding affinities yielded lower production rates because adsorption or desorption, respectively, became unfavorable in the model. This qualitative comparison supports the elemental trend captured by the kinetic model, while the transferability of the Pt-derived BEP parameters and the rate-rescaling procedure to multicomponent systems remains an approximation.

\bibliographystyle{elsarticle-num}
\bibliography{references1}